\documentclass[12pt]{iopart}

\usepackage{iopams}  
\usepackage{hyperref}
\usepackage{graphicx}
\usepackage{subcaption}
\usepackage[dvipsnames]{xcolor}

\begin{document}

\title[On superintegrability of systems with magnetic fields]{On superintegrability of 3D axially-symmetric non-subgroup-type systems with magnetic fields}

\author{S Bertrand, O Kub\r{u} and L \v{S}nobl}
\address{Czech Technical University in Prague, Faculty of Nuclear Sciences and Physical Engineering, Department of Physics, B\v{r}ehov\'a 7, 115 19 Prague 1, Czech Republic}
\ead{Sebastien.Bertrand@fjfi.cvut.cz, Ondrej.Kubu@fjfi.cvut.cz, Libor.Snobl@fjfi.cvut.cz}

\vspace{10pt}
\begin{indented}
\item\today
\end{indented}

\begin{abstract}
We extend the investigation of three-dimensional (3D) Hamiltonian systems of non-subgroup type admitting non-zero magnetic fields and an axial symmetry, namely the circular parabolic case, the oblate spheroidal case and the prolate spheroidal case. More precisely, we focus on linear and some special cases of quadratic superintegrability. In the linear case, no new superintegrable system arises. In the quadratic case, we found one new minimally superintegrable system that lies at the intersection of the circular parabolic and cylindrical cases and another one at the intersection of the cylindrical, spherical, oblate spheroidal and prolate spheroidal cases. By imposing additional conditions on these systems, we found for each quadratically minimally superintegrable system a new infinite family of higher-order maximally superintegrable systems. These two systems are linked respectively with the caged and harmonic oscillators without magnetic fields through a time-dependent canonical transformation.
\end{abstract}

\vspace{2pc}

\section{Introduction}\label{SecIntro}\setcounter{equation}{0}
The first (modern) steps in obtaining and classifying all superintegrable Hamiltonian systems were undertaken by Smorodinsky, Winternitz \textit{et al.} in \cite{FMSUW,FSUW,MSVW} concerning 2D and 3D non-relativistic Hamiltonians without magnetic fields. This research was then continued by many others, see e.g. \cite{ELW17,Evans90,Gravel04,KKM05,KKM07,KKM07o,KKM18,KWMP,MW08,MSW17,Marquette10,MPW13,TD11,VE08}. For such Hamiltonians, quadratic integrability is linked~\cite{MSVW} with the separation of variables in the Hamilton--Jacobi or Schr\"odinger equations in one of the 11 orthogonal coordinate systems listed in~\cite{EisenhartAnnMath}. However, this separability is not always preserved in the presence of a magnetic field.

Later on, a series of papers was dedicated to 2D and 3D non-relativistic Hamiltonian systems with magnetic fields for integrable and superintegrable cases, see e.g. \cite{BW04,DGRW,LMV91,MSW15,MS17,MS18,MSW18,MC10,MW00,PR05,Pucacco04,zhalij15}. One way to look for superintegrability is to start from an integrable system and then find additional integrals of motion. Up to now, only the following classes of quadratically integrable systems in 3D were investigated with magnetic fields \cite{BS19,FSW19,MS17,MSW18,zhalij15}: Cartesian, cylindrical, spherical, oblate spheroidal, prolate spheroidal and circular parabolic (also called parabolic rotational). Those classes are named after the coordinates in which the Hamilton--Jacobi / Schr\"odinger equations separate in the limit of vanishing magnetic fields. In this paper, we use the last three cases as starting points to seek for superintegrability with non-zero magnetic fields. These three integrable systems represent the systems that possess an axial symmetry and belong to the non-subgroup type of integrability, i.e. the oblate spheroidal, prolate spheroidal and circular parabolic cases. The first three classes (Cartesian, cylindrical and spherical) belong to the subgroup type and are to be investigated in other papers like \cite{MS19}. The defining property of the subgroup class is that the leading order terms of their commuting quadratic integrals are second-order Casimir operators of subgroups in some subgroup chain
\begin{eqnarray}
G\supset \tilde{G}\supset G_M,
\end{eqnarray}
where $G$ is the Euclidean group and $G_M$ is its maximal Abelian subgroup. The non-subgroup type then refers to all other classes (i.e. elliptic cylindrical, parabolic cylindrical, prolate spheroidal, oblate spheroidal, circular parabolic, parabolic, conical and ellipsoidal).

The goal of this paper is twofold. Firstly, we will complete the classification of all linearly superintegrable systems of the non-subgroup type with an axial symmetry and a magnetic field. The associated results for the circular parabolic case are already known and presented in \cite{BS19}. Secondly, we will continue the investigation of quadratic superintegrability for the circular parabolic case, the oblate spheroidal case and the prolate spheroidal case, by looking at specific (non-general) quadratic integrals of motion. The general cases are very difficult to work with because of the high numbers of constants and functions. Hence, we will focus on cases which allow separation of variables in other coordinate system(s) when there are no magnetic fields, i.e. we will follow hints from the lists of superintegrable systems in \cite{Evans90}.

The paper is structured as follows. In section \ref{SecNot}, we provide our blanket hypothesis and we specify the notation that we use throughout the paper. In section \ref{SecInt}, we summarize some results of the previous paper \cite{BS19} on integrability of the 3D non-subgroup-type Hamiltonian systems that admit non-zero magnetic fields and an axial symmetry. In section \ref{SecLin}, we classify all possible superintegrable systems admitting magnetic fields with additional integrals linear in momenta that are linked with the oblate spheroidal coordinates and the prolate spheroidal coordinates. In section \ref{SecQuadOP}, we investigate a special case of quadratic superintegrability both for the oblate spheroidal case and prolate spheroidal case, which also lies at the intersection of spherical and cylindrical cases. In section \ref{SecQuadCP}, we search for a new additional quadratic integral associated with the circular parabolic coordinates and the cylindrical coordinates. In section \ref{SecConc}, we provide some conclusions and future perspectives. Some examples of trajectories associated with minimally and maximally superintegrable systems are provided in the appendix.

\section{Notation and definitions}\label{SecNot}\setcounter{equation}{0}
We consider 3D classical Hamiltonian systems that admit a static non-vanishing magnetic field $\vec{B}$ and a static scalar potential $W$, i.e. in the Cartesian coordinates the Hamiltonian $H$ takes the form
\begin{equation}
H=\frac{1}{2}\left(\left(p_x^A\right)^2+\left(p_y^A\right)^2+\left(p_z^A\right)^2\right)+W(\vec{x}),
\end{equation}
where $p_i^A=p_i+A_i(\vec{x})$ are the covariant expressions of the momenta $p_i$ and $A_i(\vec{x})$ are the vector potential components corresponding to the magnetic field $\vec{B}(\vec{x})$. For convenience, the mass of the particle and its electric charge are set to 1 and $-1$, respectively. Throughout this paper, we prefer to use the covariant representation of the momenta to preserve the gauge invariance linked with the vector potential $\vec{A}(\vec{x})$, unless otherwise stated. Indeed, the magnetic field $\vec{B}(\vec{x})$ defines the potential vector $\vec{A}(\vec{x})$ up to a gauge transformation. In the Hamiltonian description, the choice of gauge can be seen as a canonical transformation. In our, static, case, the gauge transformation takes the form
\begin{equation}
\widetilde{A}(\vec{x})=A(\vec{x})+\nabla F(\vec{x}),\qquad \widetilde{W}(\vec{x})=W(\vec{x}).
\end{equation}
The vector potential can be interpreted as a 1-form, e.g. in the Cartesian coordinates
\begin{equation}
A=A_x(x,y,z)dx+A_y(x,y,z)dy+A_z(x,y,z)dz,
\end{equation}
such that the magnetic field 2-form $B$ is obtained by taking the exterior derivative of the vector potential $A$, e.g. in the Cartesian coordinates
\begin{eqnarray}
\hspace{-20mm}B&=&B_x(x,y,z)\, dy\wedge dz+B_y(x,y,z)\, dz\wedge dx+B_z(x,y,z)\, dx\wedge dy\nonumber\\
\hspace{-20mm}&=&(\partial_yA_z-\partial_zA_y)\, dy\wedge dz+(\partial_zA_x-\partial_xA_z)\, dz\wedge dx+(\partial_xA_y-\partial_yA_x)\, dx\wedge dy.
\end{eqnarray}
We use the notation for the angular momenta associated with the covariant linear momenta $L_x^A=yp_z^A-zp_y^A$, $L_y^A=zp_x^A-xp_z^A$ and $L_z^A=xp_y^A-yp_x^A$ together with the associated total angular momentum $(L^A)^2=(L_x^A)^2+(L_y^A)^2+(L_z^A)^2$.

\medskip

For a Hamiltonian system to be integrable (in the sense of Liouville), it must possess as many independent integrals of motion in involution as the number of dimensions. In our case, it implies that there exist 2 integrals of motion plus the Hamiltonian, where all of their Poisson brackets vanish pairwise, i.e.
\begin{equation}
\lbrace X_1,H\rbrace=0,\qquad \lbrace X_2,H\rbrace=0,\qquad \lbrace X_1,X_2\rbrace=0.
\end{equation}
The operation $\lbrace\cdot,\cdot\rbrace$ is the standard Poisson bracket. i.e.
\begin{eqnarray}
\hspace{-1cm}\lbrace f,g\rbrace=\frac{\partial f}{\partial x}\frac{\partial g}{\partial p_x}-\frac{\partial f}{\partial p_x}\frac{\partial g}{\partial x}+\frac{\partial f}{\partial y}\frac{\partial g}{\partial p_y}-\frac{\partial f}{\partial p_y}\frac{\partial g}{\partial y}+\frac{\partial f}{\partial z}\frac{\partial g}{\partial p_z}-\frac{\partial f}{\partial p_z}\frac{\partial g}{\partial z}.
\end{eqnarray}
Moreover, we assume that these integrals of motion are functionally independent, i.e. that the Jacobian matrix
\begin{equation}
\left[\frac{\partial (H,X_1,X_2)}{\partial (x_i,p_j)}\right]
\end{equation}
is of maximal rank. (Here, the maximal rank is 3.) 

If the system admits additional functionally independent integrals of motion, then it is said to be superintegrable. In the 3D case, there are only two possibilities for superintegrability: minimal superintegrability (3+1 integrals of motion) and maximal superintegrability (3+2 integrals of motion). One should note that we do not require the additional integrals of motion to be in involution with any other integral of motion except for the Hamiltonian.

In this paper, we look for additional integrals of motion that are polynomial in momenta. To obtain such additional integrals, we use a direct method, i.e. we require that the Poisson bracket of the Hamiltonian and the new integral of motion of order $N$ vanishes. The outcome of the Poisson bracket is a polynomial of order $N+1$ in momenta, for which every coefficient of the powers in momenta is zero. These coefficients are the determining equations that we need to solve to get an additional integral of motion. The determining equations are usually composed of an overdetermined system of partial differential equations involving the position coordinates. Since the cases without magnetic field are well-known in the literature, we neglect any subcases where the magnetic field vanishes. In addition, one should note that the leading-order terms of an integral of motion polynomial in momenta lie in the enveloping algebra of the Euclidean algebra $\mathfrak{e}(3)$, i.e. that they are a combination of the linear momenta $p_i$ and the angular momenta $L_i$ or, up to a redefinition of lower order terms, of their covariant versions $p_i^A$ and $L_i^A$.

More precisely, for a quadratic integral of the form
\begin{eqnarray}
X=\sum_{i<j}f_{ij}(x,y,z)p_i^Ap_j^A+\sum_{i}s_i(x,y,z)p_i^A+m(x,y,z),
\end{eqnarray}
the determining equations are given as follow:
\begin{itemize}
\item The third-order equations (implying that the quadratic terms lie in the enveloping algebra $\mathfrak{e}(3)$)
\begin{eqnarray}
\hspace{-2cm}\begin{array}{lll}
\partial_xf_{11}=0, & \partial_yf_{11}=-\partial_xf_{12}, & \partial_zf_{11}=-\partial_xf_{13},\\
\partial_xf_{22}=-\partial_yf_{12}, & \partial_yf_{22}=0, & \partial_zf_{22}=-\partial_yf_{23},\\
\partial_xf_{33}=-\partial_zf_{13},\qquad & \partial_yf_{33}=-\partial_zf_{23},\qquad & \partial_zf_{33}=0,\\
\end{array}\nonumber\\
\hspace{-18mm}\partial_xf_{23}+\partial_yf_{13}+\partial_zf_{12}=0.
\end{eqnarray}
\item The second-order equations
\begin{eqnarray}
\hspace{-23mm}\partial_xs_1=f_{13}B_y-f_{12}B_z,\quad \partial_ys_1=-\partial_xs_2-f_{13}B_x+f_{23}B_y+2(f_{11}-f_{22})B_z,\nonumber\\
\hspace{-23mm}\partial_ys_2=-f_{23}B_x+f_{12}B_z,\quad \partial_zs_2=-\partial_ys_3+2(f_{22}-f_{33})B_x-f_{12}B_y+f_{13}B_z,\\
\hspace{-23mm}\partial_zs_3=f_{23}B_x-f_{13}B_y,\quad \partial_xs_3=-\partial_zs_1+f_{12}B_x-2(f_{11}-f_{33})B_y-f_{23}B_z.\nonumber
\end{eqnarray}
\item The first-order equations
\begin{eqnarray}
\partial_xm=2f_{11}\partial_xW+f_{12}\partial_yW+f_{13}\partial_zW+s_3B_y-s_2B_z,\nonumber\\
\partial_ym=f_{12}\partial_xW+2f_{22}\partial_yW+f_{23}\partial_zW-s_3B_x+s_1B_z,\\
\partial_zm=f_{13}\partial_xW+f_{23}\partial_yW+2f_{33}\partial_zW+s_2B_x-s_1B_y.\nonumber
\end{eqnarray}
\item The zeroth-order equation
\begin{eqnarray}
s_1\partial_xW+s_2\partial_yW+s_3\partial_zW=0
\end{eqnarray}
\end{itemize}
Solving these partial differential equations is equivalent to finding a quadratic integral of motion $X$. Some additional constraints can appear in the magnetic field and the scalar potential coming from the determining equations.

When there are no magnetic fields, the equations greatly simplify. The equations split into two independent sets, one involving the functions $f_{jk}$ and $m$, the other one involving the functions $s_j$. Thus, the integral can be assumed to be either even or odd in the momenta. In \cite{MSVW}, it was proven that all 3D quadratically integrable systems are equivalent to one of the 11 systems linked with separation of coordinates and quadratic superintegrability was addressed for the systems separating in the spherical coordinates. Evans in his PhD thesis extended this study to all quadratically superintegrable systems in 3D. He has shown that these systems separate in more than one coordinate set and thus lie at the intersection of integrable classes. In~\cite{Evans90} he provided a list of all 3D quadratically minimally and maximally superintegrable Hamiltonian systems (non-relativistic, without magnetic fields and time-independent) together with the coordinate systems in which they separate and the corresponding integrals of motion.

\section{Past results on integrability for axially-symmetric Hamiltonians}\label{SecInt}\setcounter{equation}{0}
In this paper, we consider 3D non-subgroup-type integrable systems with non-zero magnetic fields and an axial symmetry as starting points. These systems were previously studied in \cite{BS19}, and are linked with three types of coordinates:
\begin{itemize}
\item the circular parabolic coordinates,
\item the oblate spheroidal coordinates,
\item the prolate spheroidal coordinates.
\end{itemize}
The subgroup-type integrable systems with non-zero magnetic fields and an axial symmetry (cylindrical and spherical cases) were treated in other papers \cite{FSW19,MSW18}. These subgroup-type integrable systems will not be used as starting points here, however, the superintegrable systems found in this paper lie at the intersection of subgroup-type and non-subgroup-type cases.

To make the paper self-contained, we briefly provide results concerning the integrability of non-subgroup-type integrable systems with non-zero magnetic fields and an axial symmetry.

\subsection{The circular parabolic integrable case}\label{SSecCP}
The circular parabolic coordinates are given through their transformation into the Cartesian coordinates as
\begin{equation}
x=\xi\eta\cos(\phi),\qquad y=\xi\eta\sin(\phi),\qquad z=\frac{1}{2}(\xi^2-\eta^2).
\end{equation}
This system of coordinates is alternatively called the parabolic rotational coordinates. The integrable Hamiltonian associated with the circular parabolic case is 
\begin{eqnarray}
\hspace{-20mm}H&=&\frac{1}{2}\left(\frac{\left(p_\xi^A\right)^2+\left(p_\eta^A\right)^2}{\xi^2+\eta^2}+\frac{\left(p_\phi^A\right)^2}{\xi^2\eta^2}\right)+\frac{1}{2}\left(\frac{\beta_1(\eta)-\beta_2(\xi)}{\xi^2+\eta^2}\right)^2+\frac{\eta^2\rho_2(\xi)-\xi^2\rho_1(\eta)}{\xi^2\eta^2(\xi^2+\eta^2)}
\end{eqnarray}
together with the magnetic field
\begin{eqnarray}
\hspace{-20mm}B_\xi=-\partial_\eta\left(\frac{\xi^2\beta_1(\eta)+\eta^2\beta_2(\xi)}{\xi^2+\eta^2}\right),\quad B_\eta=\partial_\xi\left(\frac{\xi^2\beta_1(\eta)+\eta^2\beta_2(\xi)}{\xi^2+\eta^2}\right),\quad B_\phi=0,
\end{eqnarray}
where $\rho_1(\eta)$ and $\rho_2(\xi)$ are arbitrary functions appearing solely in the scalar potential, while $\beta_1(\eta)$ and $\beta_2(\xi)$ are arbitrary functions that appear also in the magnetic field. This system possesses two quadratic integrals of motion: a (Laplace--)Runge--Lenz-type integral of motion
\begin{eqnarray}
\hspace{-20mm}X_1&=&L_x^Ap_y^A-L_y^Ap_x^A+\mbox{lower order terms}\\
\hspace{-20mm}&=&\frac{\eta^2(p_\xi^A)^2-\xi^2(p_\eta^A)^2}{2(\xi^2+\eta^2)}+\frac{\eta^2-\xi^2}{2\xi^2\eta^2}(p_\phi^A)^2+\left(\frac{\beta_1(\eta)-\beta_2(\xi)}{\xi^2+\eta^2}\right)p_\phi^A+\frac{\xi^4\rho_1(\eta)+\eta^4\rho_2(\xi)}{\xi^2\eta^2(\xi^2+\eta^2)}\nonumber
\end{eqnarray}
and a quadratic angular momentum integral of motion that degenerates in the presence of a magnetic field to a linear one, namely
\begin{eqnarray}
X_2&=&L_z^A+\mbox{lower order terms}\\
&=&p_\phi^A+\frac{\xi^2\beta_1(\eta)+\eta^2\beta_2(\xi)}{\xi^2+\eta^2},\nonumber
\end{eqnarray}
corresponding to the axial symmetry. The integral of motion $X_2$ becomes simply $p_\phi$ by using the suitable choice of gauge for the vector potential $A$
\begin{eqnarray}
A_\xi=A_\eta=0,\qquad A_\phi=-\frac{\xi^2\beta_1(\eta)+\eta^2\beta_2(\xi)}{\xi^2+\eta^2}.
\end{eqnarray}

\subsection{The oblate spheroidal integrable case}\label{SSecOS}
The oblate spheroidal coordinates are given through their transformation into the Cartesian coordinates as
\begin{eqnarray}
x&=&a\cosh(\xi)\sin(\eta)\cos(\phi),\qquad y=a\cosh(\xi)\sin(\eta)\sin(\phi),\label{transob}\\
z&=&a\sinh(\xi)\cos(\eta),\nonumber
\end{eqnarray}
where $a$ is a parameter greater than zero. The integrable Hamiltonian associated with the oblate spheroidal case is 
\begin{eqnarray}
H&=&\frac{1}{2}\left(\frac{(p_\xi^A)^2+(p_\eta^A)^2}{a^2(\cosh^2(\xi)-\sin^2(\eta))}+\frac{(p_\phi^A)^2}{a^2\cosh^2(\xi)\sin^2(\eta)}\right)\nonumber\\
& &-\frac{1}{2}\left(\frac{\beta_1(\eta)-\beta_2(\xi)}{2a(\cosh^2(\xi)-\sin^2(\eta))}\right)^2+\frac{\rho_1(\eta)+\rho_2(\xi)}{2a^2(\cosh^2(\xi)-\sin^2(\eta))}
\end{eqnarray}
together with the magnetic field
\begin{eqnarray}
B_\xi &=&\partial_\eta\left(\frac{\sin^2(\eta)\beta_2(\xi)-\cosh^2(\xi)\beta_1(\eta)}{2(\cosh^2(\xi)-\sin^2(\eta))}\right),\nonumber\\
B_\eta &=&-\partial_\xi\left(\frac{\sin^2(\eta)\beta_2(\xi)-\cosh^2(\xi)\beta_1(\eta)}{2(\cosh^2(\xi)-\sin^2(\eta))}\right),\\
B_\phi &=&0,\nonumber
\end{eqnarray}
where $\rho_1(\eta)$ and $\rho_2(\xi)$ are arbitrary functions appearing solely in the scalar potential, while $\beta_1(\eta)$ and $\beta_2(\xi)$ are arbitrary functions that appear also in the magnetic field. This system possesses two quadratic integrals of motion: the integral of motion
\begin{eqnarray}
\hspace{-1cm}X_1&=&(L^A)^2+a^2\left((p_x^A)^2+(p_y^A)^2\right)+\mbox{lower order terms}\label{rawX1OS}\\
&=&\frac{\sin^2(\eta)(p_\xi^A)^2+\cosh^2(\xi)(p_\eta^A)^2}{\cosh^2(\xi)-\sin^2(\eta)}+\frac{\cosh^2(\xi)+\sin^2(\eta)}{\cosh^2(\xi)\sin^2(\eta)}(p_\phi^A)^2\nonumber\\
& &+\left(\frac{\beta_1(\eta)-\beta_2(\xi)}{\cosh^2(\xi)-\sin^2(\eta)}\right)p_\phi^A+\frac{\cosh^2(\xi)\rho_1(\eta)+\sin^2(\eta)\rho_2(\xi)}{\cosh^2(\xi)-\sin^2(\eta)},\nonumber
\end{eqnarray}
and a quadratic angular momentum integral of motion that degenerates to a linear one, i.e.
\begin{eqnarray}
X_2&=&L_z^A+\mbox{lower order terms}\label{rawX2OS}\\
&=&p_\phi^A+\frac{\cosh^2(\xi)\beta_1(\eta)-\sin^2(\eta)\beta_2(\xi)}{2(\cosh^2(\xi)-\sin^2(\eta))},\nonumber
\end{eqnarray}
corresponding to the axial symmetry. The integral of motion $X_2$ becomes simply $p_\phi$ by using the suitable choice of gauge for the vector potential $A$
\begin{eqnarray}
A_\xi=A_\eta=0,\qquad A_\phi=-\frac{\cosh^2(\xi)\beta_1(\eta)-\sin^2(\eta)\beta_2(\xi)}{2(\cosh^2(\xi)-\sin^2(\eta))}.
\end{eqnarray}

\subsection{The prolate spheroidal integrable case}\label{SSecPS}
The prolate spheroidal coordinates are given through their transformation into the Cartesian coordinates as
\begin{eqnarray}
x&=&a\sinh(\xi)\sin(\eta)\cos(\phi),\qquad y=a\sinh(\xi)\sin(\eta)\sin(\phi),\label{transpro}\\
z&=&a\cosh(\xi)\cos(\eta),\nonumber
\end{eqnarray}
where $a$ is a parameter greater than zero. The integrable Hamiltonian associated with the prolate spheroidal case is
\begin{eqnarray}
H&=&\frac{1}{2}\left(\frac{(p_\xi^A)^2+(p_\eta^A)^2}{a^2(\sinh^2(\xi)+\sin^2(\eta))}+\frac{(p_\phi^A)^2}{a^2\sinh^2(\xi)\sin^2(\eta)}\right)\nonumber\\
& &+\frac{1}{2}\left(\frac{\beta_1(\eta)-\beta_2(\xi)}{2a(\sinh^2(\xi)+\sin^2(\eta))}\right)^2+\frac{\rho_1(\eta)+\rho_2(\xi)}{2a^2(\sinh^2(\xi)+\sin^2(\eta))}
\end{eqnarray}
together with the magnetic field
\begin{eqnarray}
B_\xi &=&-\partial_\eta\left(\frac{\sin^2(\eta)\beta_2(\xi)+\sinh^2(\xi)\beta_1(\eta)}{2(\sinh^2(\xi)+\sin^2(\eta))}\right),\nonumber\\
B_\eta &=&\partial_\xi\left(\frac{\sin^2(\eta)\beta_2(\xi)+\sinh^2(\xi)\beta_1(\eta)}{2(\sinh^2(\xi)+\sin^2(\eta))}\right),\\
B_\phi &=&0,\nonumber
\end{eqnarray}
where $\rho_1(\eta)$ and $\rho_2(\xi)$ are arbitrary functions appearing solely in the scalar potential, while $\beta_1(\eta)$ and $\beta_2(\xi)$ are arbitrary functions that appear also in the magnetic field. This system possesses two quadratic integrals of motion: the integral of motion
\begin{eqnarray}
\hspace{-1cm}X_1&=&(L^A)^2-a^2\left((p_x^A)^2+(p_y^A)^2\right)+\mbox{lower order terms}\label{rawX1PS}\\
&=&\frac{\sinh^2(\xi)(p_\eta^A)^2-\sin^2(\eta)(p_\xi^A)^2}{\sinh^2(\xi)+\sin^2(\eta)}+\frac{\sinh^2(\xi)-\sin^2(\eta)}{\sinh^2(\xi)\sin^2(\eta)}(p_\phi^A)^2\nonumber\\
& &+\left(\frac{\beta_2(\xi)-\beta_1(\eta)}{\sinh^2(\xi)+\sin^2(\eta)}\right)p_\phi^A+\frac{\sinh^2(\xi)\rho_1(\eta)-\sin^2(\eta)\rho_2(\xi)}{\sinh^2(\xi)+\sin^2(\eta)}\nonumber
\end{eqnarray}
and a quadratic integral of motion that degenerates to a linear one, i.e.
\begin{eqnarray}
X_2&=&L_z^A+\mbox{lower order terms}\label{rawX2PS}\\
&=&p_\phi^A+\frac{\sinh^2(\xi)\beta_1(\eta)+\sin^2(\eta)\beta_2(\xi)}{2(\sinh^2(\xi)+\sin^2(\eta))},\nonumber
\end{eqnarray}
corresponding to the axial symmetry. The integral of motion $X_2$ becomes simply $p_\phi$ by using the suitable choice of gauge for the vector potential $A$
\begin{eqnarray}
A_\xi=A_\eta=0,\qquad A_\phi=-\frac{\sinh^2(\xi)\beta_1(\eta)+\sin^2(\eta)\beta_2(\xi)}{2(\sinh^2(\xi)+\sin^2(\eta))}.
\end{eqnarray}

\section{Linear superintegrability: oblate and prolate spheroidal}\label{SecLin}\setcounter{equation}{0}
In this section, we investigate the linear superintegrability associated with the oblate spheroidal case and the prolate spheroidal case. We look separately into the oblate and prolate spheroidal cases for a general additional linear integral, which takes the form
\begin{equation}
Y_3=\alpha_1p_x^A+\alpha_2p_y^A+\alpha_3p_z^A+\alpha_4L_x^A+\alpha_5L_y^A+\alpha_6L_z^A+m(x,y,z).
\end{equation}
The constant $\alpha_6$ can be set to zero in both the oblate and prolate spheroidal cases since the new integral must be functionally independent with the integral $X_2$. After doing such an investigation for both the oblate and prolate spheroidal cases, taking out the cases where the magnetic fields vanish, we are left with two superintegrable systems that appear for both the oblate spheroidal case and the prolate spheroidal case.

The first superintegrable system involves the additional integral of motion
\begin{equation}
Y_3=p_z^A,\label{Sec42}
\end{equation}
which leads to the Hamiltonian
\begin{equation}
H=\frac{(p_x^A)^2+(p_y^A)^2+(p_z^A)^2}{2}+\frac{u_1}{x^2+y^2}-\frac{b_z^2}{8}(x^2+y^2)\label{Sec43}
\end{equation}
together with the constant magnetic field oriented along the $z$-axis
\begin{equation}\label{konstB}
B=b_z dx\wedge dy.
\end{equation}
The corresponding vector potential can be chosen in the form
\begin{eqnarray}\label{AkonstB}
A=-\frac{b_z}{2}y dx+\frac{b_z}{2}x dy.
\end{eqnarray}
The integrals of motion (\ref{rawX1OS}-\ref{rawX2OS}) and (\ref{rawX1PS}-\ref{rawX2PS}) written in the Cartesian coordinates become
\begin{eqnarray}
\hspace{-10mm}X_1&=&(L^A)^2-b_z(x^2+y^2+z^2)L_z^A+\frac{2u_1z^2}{x^2+y^2}+\frac{b_z^2}{4}(x^2+y^2)(x^2+y^2+z^2),\label{Sec44}\\
\hspace{-10mm}X_2&=&L_z^A-\frac{b_z}{2}(x^2+y^2).\label{Sec45}
\end{eqnarray}
The integral $X_1$ has been redefined using the other integrals of motion to get rid of the parameter $a$ appearing in the oblate / prolate coordinates definition (\ref{transob}) / (\ref{transpro}). The linear and quadratic integrals of motion (\ref{Sec42}-\ref{Sec43}) and (\ref{Sec44}-\ref{Sec45}) are functionally independent,  thus the system is minimally superintegrable.

The second superintegrable system involves two independent additional integrals of motion,
\begin{equation}
Y_3=p_x^A+b_zy,\qquad Y_4=p_y^A-b_zx,
\end{equation}
which leads to a Hamiltonian similar to (\ref{Sec43}) but with $(x^2+y^2)$ replaced by $z^2$ and one sign flipped,
\begin{equation}
H=\frac{(p_x^A)^2+(p_y^A)^2+(p_z^A)^2}{2}+\frac{u_1}{z^2}+\frac{b_z^2}{8}z^2,
\end{equation}
with the same constant magnetic field~(\ref{konstB}) oriented along the $z$-axis
\begin{equation}
B=b_z dx\wedge dy.
\end{equation}
(Thus, the vector potential reads as in~(\ref{AkonstB}).) The integrals of motion (\ref{rawX1OS}-\ref{rawX2OS}) and (\ref{rawX1PS}-\ref{rawX2PS}) written in the Cartesian coordinates are
\begin{eqnarray}
\hspace{-20mm}X_1&=&(L^A)^2-b_z(x^2+y^2+z^2)L_z^A+\frac{2u_1}{z^2}(x^2+y^2)+\frac{b_z^{2}}{4}(x^2+y^2)(x^2+y^2+z^2),\\
\hspace{-20mm}X_2&=&L_z^A-\frac{b_z}{2}(x^2+y^2).
\end{eqnarray}
Similarly, the integral $X_1$ has been redefined to get rid of the parameter $a$. This system possesses 5 functionally independent integrals of motion, i.e. it  is maximally superintegrable.

These two systems are already known in the literature, see e.g. \cite{BS19,MS17,MS19}. One should note that the classification of all linearly superintegrable systems for the circular parabolic case was provided in section 6 of the paper \cite{BS19}.

\pagebreak

\section{Special quadratic superintegrability: oblate and prolate spheroidal}\label{SecQuadOP}\setcounter{equation}{0}
In this section and the next one, we do not look for general quadratic integrals of motion. The number of leading order constants together with the number of functions depending on position coordinates becomes too difficult to manage without giving a blind control to computer algebra systems like Maple \cite{Maple}. It gets drastically worse every time the order of a general integral of motion is raised. Even if computer algebra systems like Maple are very useful and powerful, they are not perfect when we need to solve partial differential equations and some solutions may be missed. Therefore, we will be looking at special cases which would allow us to search for quadratic superintegrability without relying on the power of symbolic solvers of partial differential equations. More precisely, we consider cases that remain quadratically superintegrable when the magnetic field is set to zero, and thus according to the results of \cite{Evans90,MSVW} allow separation of variables in one of the sets of coordinates under study and also in at least another set of coordinates. In other words, we consider second-order integrals of motion that, when the magnetic fields are set to zero, would lead to the separation of variables in the oblate spheroidal, prolate spheroidal or circular parabolic coordinates and in another coordinate system. Evans \cite{Evans90} provided such a list of all quadratically minimally and maximally superintegrable systems when there is no magnetic field. Assuming that a superintegrable system with a magnetic field has a meaningful nontrivial limit as the magnetic field goes to zero, we may reasonably expect that the integrals turn into integrals found by Evans in the limit of vanishing magnetic fields. Since the leading order determining equations do not involve the magnetic field, we may impose the assumption that the leading order terms are not affected by the limit. Thus, starting with the same leading order terms of additional integrals, we look for additional quadratic integrals of motion when the magnetic field does not vanish.

According to Evans \cite{Evans90}, when there are no magnetic fields, there exist one quadratically minimally superintegrable system involving the separation of variables in the oblate and prolate spheroidal coordinates and one quadratically maximally superintegrable system involving separation of variables in the oblate and prolate spheroidal coordinates, see the fifth case in Table II and the first case in Table I of \cite{Evans90}, respectively. By considering the system with a magnetic field corresponding to the maximally superintegrable case without a magnetic field, the additional quadratic integrals lead to a maximally superintegrable system that possesses many linear integrals of motion in terms of which the imposed quadratic ones are expressed as their functions. This system has already been found in the previous section and e.g. in \cite{BS19,MS17, MS19}. 

Hence, we look at the minimally superintegrable case, i.e. the overlap with the cylindrical and spherical cases, that is we impose an additional integral of the form
\begin{equation}
Y_3=(p_z^A)^2+\mbox{lower order terms},\label{rawY3OP}
\end{equation}
or the same investigation could have been done using the additional integral
\begin{equation}
(L^A)^2+\mbox{lower order terms}.
\end{equation}
That comes from the fact that it is possible to use the Hamiltonian and (\ref{rawY3OP}) to get rid of the parameter $a$ in this system, i.e.
\begin{equation}
(L^A)^2+...=X_1\pm a^2Y_3 \mp 2a^2H.
\end{equation}

We searched for the integral (\ref{rawY3OP}) for the oblate spheroidal case and the prolate spheroidal case, separately. From the second-order determining equations of the Poisson brackets of the three integrals $X_1$, $X_2$ and $Y_3$ with the Hamiltonian $H$, we can take the compatibility conditions for the linear terms of the integrals of motion to get an overdetermined system of linear partial differential equations for the magnetic field, which can be solved. Once the admissible magnetic field is found from the compatibility conditions, one can solve the second-order determining equations to get the linear coefficients in the momenta. Then, by considering the compatibility conditions of the linear determining equations together with the zeroth-order determining equations, the potential $W$ can be found. As expected, we found the same system for the oblate and prolate spheroidal cases (and the spherical case) after removing the non-interesting cases (vanishing magnetic field and systems which reduce to the already known ones with linear integrals).  The new Hamiltonian system is
\begin{eqnarray}
H&=&\frac{(p_x^A)^2+(p_y^A)^2+(p_z^A)^2}{2}+\frac{u_1}{r^2}+\frac{u_2}{z^2}-u_3R^2-\frac{b_pb_s}{4z^2}R^4\nonumber\\
& &-\frac{b_zb_p}{4z^2}r^2-\frac{b_zb_s}{4}r^2R^2-\frac{b_s^2}{8}r^2R^4+\frac{b_z^2}{8}z^2-\frac{b_p^2}{8z^4}r^2\label{OPHmin}
\end{eqnarray}
together with the magnetic field
\begin{eqnarray}
B&=&\left(\frac{b_px}{z^3}-b_sxz\right)dy\wedge dz+\left(\frac{b_py}{z^3}-b_syz\right)dz\wedge dx\nonumber\\
& &+\left(b_z+\frac{b_p}{z^2}+b_s(r^2+R^2)\right)dx\wedge dy,
\end{eqnarray}
where $r$ is the cylindrical radius, i.e. $r^2=x^2+y^2$, and $R$ is the spherical radius, i.e. $R^2=x^2+y^2+z^2$. The constants $u_i$ appear only in the scalar potential while the constants $b_i$ also appear in the magnetic field. The  vector potential $A$ can be chosen as
\begin{equation}
A= -\frac{y}{2}\left(b_z+\frac{b_p}{z^2}+b_sR^2\right) dx+\frac{x}{2}\left(b_z+\frac{b_p}{z^2}+b_sR^2\right) dy.
\end{equation}
The integrals of motion are given by
\begin{eqnarray}
X_1&=&(L^A)^2-(b_z+b_sR^2)R^2L_z^A+\frac{2u_1}{r^2}z^2+\frac{2u_2}{z^2}r^2\nonumber\\
& & +\frac{b_z^2}{4}r^2R^2+\frac{b_zb_s}{2}r^2R^4-\frac{b_p^2}{4z^4}r^2R^2+\frac{b_s^2}{4}r^2R^6,\\
X_2&=&L_z^A-\frac{b_z}{2}r^2-\frac{b_p}{2z^2}r^2-\frac{b_s}{2}r^2R^2,\\
Y_3&=&(p_z^A)^2+\left(\frac{b_p}{z^2}+b_sz^2\right)L_z^A+\frac{2u_2}{z^2}-2u_3z^2+\frac{b_z^2}{4}z^2\nonumber\\
& &-\frac{b_zb_p}{2z^2}r^2-\frac{b_zb_s}{2}z^2r^2-\frac{b_p^2}{2z^4}r^2-\frac{b_pb_s}{2z^2}R^4-\frac{b_s^2}{2}z^2r^2R^2.
\end{eqnarray}
The Poisson bracket $\lbrace X_1,Y_3\rbrace$ is the only one that is not zero, but the algebra closes polynomially, i.e. the Poisson bracket squared can be expressed as a polynomial in terms of $H$, $X_1$, $X_2$ and $Y_3$,
\begin{eqnarray}
\hspace{-20mm}\left(\lbrace X_1,Y_3\hspace{-3mm}\phantom{^{2^2}}\rbrace\right)^2&=& 32 H X_1 Y_3 - 32 H X_2^2 Y_3 - 16 X_1 Y_3^2+ 16X_2 (-b_s X_2^4  \nonumber\\
& &  + 2b_s X_1 X_2^2 + b_z X_2^2 Y_3  - 4 b_p H^2  + 2b_p H Y_3  - b_s X_1^2\nonumber\\
& &  - b_z X_1 Y_3 )- 128 u_2 H^2 + 64 b_p b_z H X_2^2 + 128 u_2 H Y_3\nonumber\\
& & + 4( 2 b_p b_s- b_z^2 + 8 u_3)X_1^2  + 8( b_z^2 + 2 b_p b_s - 8 u_3)X_1 X_2^2 \nonumber\\
& & + 4(10 b_p b_s + 8 u_3- b_z^2 )X_2^4  - 16 b_p b_z X_2^2 Y_3\nonumber\\
& & -32 (u_1 + u_2)Y_3^2  + 8 (16 b_z u_2 H+b_p ( b_z^2 - 2 b_p b_s - 8 u_3)X_1  \nonumber\\
& & + (16 b_s u_2- b_z^2 b_p - 4 b_p^2 b_s - 8 b_p u_3 )X_2^2  - 8 b_z u_2 Y_3)X_2 \nonumber\\
& & + 4(2 b_p^3 b_s- b_z^2 b_p^2 + 8 b_p^2 u_3 + 32 b_p b_s u_1 - 16 b_p b_s u_2 - 64 u_2 u_3)X_2^2 \nonumber\\
& & + 32 u_1 (b_z^2 b_p - 2 b_p^2 b_s - 8 b_p u_3 + 8 b_s u_2)X_2 \nonumber\\
& & + 64 u_1 u_2 (b_z^2 - 2 b_p b_s - 8 u_3).
\end{eqnarray}
It is interesting to observe that the involutions of $X_1$ and $Y_3$ with $X_2$ are in this case obtained as a consequence of the existence of the integrals, i.e. there was no need to assume or impose their involution a priori.

We can solve the associated equations of motion in the cylindrical coordinates,
\begin{eqnarray}
\dot{r}=p_r,\qquad \dot{p_r}= -\left(b_sp_\theta - 2u_3+\frac{b_z^2}{4}-\frac{b_pb_s}{2}\right)r+\frac{p_\theta^2 + 2u_1}{r^3},\nonumber\\
\dot{z}=p_z,\qquad \dot{p_z}=-\left(b_sp_\theta - 2u_3+\frac{b_z^2}{4}-\frac{b_pb_s}{2}\right)z + \frac{ b_pp_\theta + 2u_2}{z^3},\label{eqmothetaOP}\\
\dot{\theta}=\frac{b_s}{2}r^2+\frac{p_\theta}{r^2}+\frac{b_s}{2}z^2+\frac{b_p}{2z^2}+\frac{b_z}{2},\qquad \dot{p_\theta}=0.\nonumber
\end{eqnarray}
The solution takes the form
\begin{eqnarray}
r(t)&=&\sqrt{c_1\cos(\nu t+c_2)+\sqrt{c_1^2+4\frac{L_z^2+2u_1}{\nu^2}}},\label{solrOP}\\
z(t)&=&\epsilon\sqrt{c_3\cos(\nu t+c_4)+\sqrt{c_3^2+4\frac{b_pL_z+2u_2}{\nu^2}}},\qquad \epsilon^2=1,\label{solzOP}\\
\theta(t)&=&c_5+k_1t+\frac{b_s}{2\nu}\left(c_1\sin(\nu t+c_2)+c_3\sin(\nu t+c_4)\right)\nonumber\\
& &+\frac{L_z}{\sqrt{L_z^2+2u_1}}\arctan\left(k_2\tan\left(\frac{\nu t+c_2}{2}\right)\right)\nonumber\\
& &+\frac{b_p}{2\sqrt{b_pL_z+2u_2}}\arctan\left(k_3\tan\left(\frac{\nu t+c_4}{2}\right)\right),\label{solthetaOP}
\end{eqnarray}
where the $c_i$ and $L_z$ are integration constants and the constants $k_i$ are given by 
\begin{eqnarray}
k_1&=&\frac{b_z}{2}+\frac{b_s}{2}\left(\sqrt{c_1^2+4\frac{L_z^2+2u_1}{\nu^2}}+\sqrt{c_3^2+4\frac{L_zb_p+2u_2}{\nu^2}}\right),\\
k_2&=&\sqrt{1+\frac{c_1^2\nu^2}{4(L_z^2+2u_1)}}-\frac{c_1\nu}{2\sqrt{L_z^2+2u_1}},\\
k_3&=&\sqrt{1+\frac{c_3^2\nu^2}{4(L_zb_p+2u_2)}}-\frac{c_3\nu}{2\sqrt{L_z b_p+2u_2}}.
\end{eqnarray}
The frequency $\nu$ of $r$ and $z$ is determined by the (initial) angular momentum $L_z$ and the constants $b_z$, $b_p$, $b_s$ and $u_3$, i.e.
\begin{equation}
\nu=\sqrt{b_z^2-8u_3+2b_s(2L_z-b_p)}.\label{freq1}
\end{equation}

We see that this system does not possess the behaviour of a maximally superintegrable system \cite{Nehorosev}, i.e. its bounded trajectories are not closed, unless there are some additional restrictions. Hence, we conclude that the system is only minimally superintegrable in its general form. However, particular superintegrability in the sense of \cite{Turbiner} can appear when $\theta$ matches the frequency of $r$ and $z$ up to a multiplication by a rational number.

When $b_s=b_p=0$, the frequency $\nu$ of $r$ and $z$ becomes independent of the initial conditions (namely the conserved angular momentum $L_z$), that is
\begin{eqnarray}
\nu=\sqrt{b_z^2-8u_3}.
\end{eqnarray}
The remaining magnetic field is constant and oriented along the $z$-axis. By imposing the additional constraint
\begin{eqnarray}
u_3=\frac{b_z^2}{8}\left(1-\frac{n^2}{4m^2}\right),\qquad n,m\in\mathbb{N}\label{cond5}
\end{eqnarray}
the frequency of $r$ and $z$ will match the frequency generated by $\theta$ up to the multiplication by some integers $n$ and $m$. This relation is obtained by matching the frequency $\nu$ with the constant $k_1$. (One should remember that $\theta$ is an angular variable, hence the constant $k_1$ corresponds to a frequency in this case.) Under these conditions, the system becomes isochronous \cite{Calogero08}. Using the rotating-frame transformation around the $z$-axis \cite{RotF}
\begin{eqnarray}
R(t)=\left(\begin{array}{ccc}
\cos(\frac{b_z}{2}t) & -\sin(\frac{b_z}{2}t) & 0\\
\sin(\frac{b_z}{2}t) & \cos(\frac{b_z}{2}t) & 0 \\
 0 & 0 &1
\end{array}\right)\label{rotFrame5}
\end{eqnarray}
we can get rid of the magnetic field $b_z$. Under the additional constraint $u_1=0$, we obtain a caged oscillator
\begin{eqnarray}\label{cagedosc}
H=\frac{p_x^2+p_y^2+p_z^2}{2}+\frac{u_2}{z^2}+\frac{\nu^2}{8}(x^2+y^2+z^2),\qquad \nu=\frac{nb_z}{2m},
\end{eqnarray}
without a magnetic field, which is known to be maximally superintegrable, cf. \cite{Evans90,EV08}. If $u_1$ is not set to zero, Bertrand's theorem applied to the $xy$--plane implies that the trajectories of the magnetic-less rotated system cannot close, hence the system cannot be maximally superintegrable. In that case, the procedure below does not yield enough integrals for the original system.

We can map back the integrals of motion of the system~(\ref{cagedosc}) (one of which we obtain from the ladder operator following \cite{Marquette12}) and eliminate the time dependence arising through the rotation by taking a suitable combination of them analogously to \cite{EV08} -- that is possible under the rationality condition (\ref{cond5}) relating $u_3$ and $b_z$. Thus, we obtain a maximally superintegrable system. Its Hamiltonian reads
\begin{eqnarray}
H=\frac{(p_x^A)^2+(p_y^A)^2+(p_z^A)^2}{2}+\frac{u_2}{z^2}-\frac{b_z^2}{8}r^2+\frac{n^2b_z^2}{32m^2}R^2\label{MaxHam5}
\end{eqnarray}
with a constant magnetic field of magnitude $b_z$ oriented along the $z$-axis. The integrals of motion are
\begin{eqnarray}
X_1&=&(L^A)^2-b_zR^2L_z^A+\frac{2u_2}{z^2}r^2+\frac{b_z^2}{4}r^2R^2,\\
X_2&=&L_z^A-\frac{b_z}{2}r^2,\\
Y_3&=&(p_z^A)^2+\frac{2u_2}{z^2}+\frac{n^2b_z^2}{16m^2}z^2,\\
Y_4&=&\mbox{Re}\left(\left(\frac{8{\rm i}mnb_zz^3p_z^A + b_z^2n^2z^4 - 16m^2z^2(p_z^A)^2 - 32m^2u_2}{nb_zz^2}\right)^{2m}\right.\nonumber\\
& &\times\left(4b_z^2(y- {\rm i}x)^2 + 16b_z(x+{\rm i}y)(p_y^A-{\rm i}p_x^A)\right.\nonumber\\
& &\left.\left.+16\left(p_x^A+{\rm i}p_y^A\right)^2+\frac{n^2b_z^2(x+{\rm i}y)^2}{m^2}\right)^n\right)
\end{eqnarray}
The integral $Y_4$ is of order $2 (n +2m)$ in momenta and can be expressed explicitly without using complex expressions in terms of Chebyshev polynomials as in \cite{EV08} or in \cite{MS18}, cf. equations (18-20) and (3.7) therein, respectively. As an example, in the case $m=n=1$, the integral $Y_4$ becomes up to a rescaling by a numerical constant
\begin{eqnarray}
\hspace{-10mm}Y_4&=&((p_x^A)^2-(p_y^A)^2)(p_z^A)^4+b_z(p_z^A)^3(2 z p_x^A p_y^A + y p_x^A p_z^A + x p_y^A p_z^A)\nonumber\\
\hspace{-10mm}& &-\frac{3(p_z^A)^2}{16 z^2} \left(2b_z^2z^4((p_x^A)^2 - (p_y^A)^2) + \frac{16b_z^2}{3}z^3 (x p_x^A - y p_y^A)p_z^A\right.\nonumber\\
\hspace{-10mm}& & \left.+ b_z^2 (x^2 - y^2)z^2(p_z^A)^2 - \frac{64 u_2}{3} \left((p_x^A)^2 - (p_y^A)^2\right)\right)\nonumber\\
\hspace{-10mm}& & -\frac{3b_z p_z^A}{8z^2}\left(b_z^2 x y z^3 (p_z^A)^2 +\frac{3b_z^2z^4 - 32u_2}{3} (y p_x^A + x p_y^A)p_z^A\right.\nonumber\\
\hspace{-10mm}& & \left.+\frac{b_z^2 z^4 - 32u_2}{3} z p_x^A p_y^A\right)+\frac{(b_z^2z^4 - 32u_2)^2}{256z^4}(p_x^A)^2\nonumber\\
\hspace{-10mm}& & +b_z^2\frac{b_z^2 z^4-32 u_2}{16z}x p_x^A p_z^A-\frac{(b_z^2z^4 - 32u_2)^2}{256z^4}(p_y^A)^2-b_z^2\frac{b_z^2 z^4 - 32 u_2}{16z} y p_y^A p_z^A\nonumber\\
\hspace{-10mm}& & +\frac{3 b_z^2}{128 z^2} (3 b_z^2 z^4-32 u_2)  (x^2-y^2) (p_z^A)^2+\frac{3b_z}{128z^4}(b_z^2 z^4 - 32 u_2)\nonumber\\
\hspace{-10mm}& & \times\left(\frac{b_z^2}{6}z^4(y p_x^A + xp_y^A) + b_z^2 x y z^3 p_z^A -\frac{16 u_2}{3} (y p_x^A + x p_y^A)\right)\nonumber\\
\hspace{-10mm}& & -\frac{3 b_z^2}{4096 z^4}(b_z^2 z^4 - 32 u_2)^2 (x^2 - y^2).
\end{eqnarray}

Let us remark that this construction of maximally superintegrable systems is not exhaustive in the sense that there may exist other maximally superintegrable systems of the form~(\ref{OPHmin}) for some other special values of its parameters. We used the system~(\ref{cagedosc}) without magnetic field to generate a previously unknown integral for the system~(\ref{MaxHam5}) with magnetic field. This procedure, however, doesn't exclude a hypothetical existence of systems with magnetic field whose trajetories close but are not isoperiodic with the rotating frame transformation~(\ref{rotFrame5}), i.e. the trajectories of the corresponding magnetic-less systems no longer close and thus are not maximally superintegrable. Hence such maximally superintegrable systems with magnetic field would be impossible to recover via our method. In addition, we focused on isochronous systems, see e.g. \cite{Calogero08,GL18}, i.e. maximally superintegrable systems with a frequency independent of the initial value.

\section{Special quadratic superintegrability: circular parabolic}\label{SecQuadCP}\setcounter{equation}{0}
According to Evans \cite{Evans90}, when there are no magnetic fields, there exist 2 quadratically minimally superintegrable systems allowing separation of variables in the circular parabolic coordinates and 3 maximally superintegrable ones, see the sixth and seventh cases in Table II and the second, fourth and fifth cases in Table I of \cite{Evans90}, respectively. By considering the systems with magnetic fields corresponding to the ``maximally'' superintegrable cases without magnetic fields, the additional quadratic integrals with the corresponding leading order terms lead again to maximally superintegrable systems that possess many linear integrals of motion. The quadratic integrals can be expressed in terms of the linear ones. These systems have already been found in e.g. \cite{BS19,MS17, MS19}. For the minimally superintegrable counterparts, the overlap with the spherical case was studied in \cite{BS19}, where we found a new quadratically superintegrable system. Details on this system can be found in section 7 of \cite{BS19}.

However, the overlap with the cylindrical case was not studied previously, i.e. looking for an additional integral of motion of the form
\begin{equation}
Y_3=(p_z^A)^2+\mbox{lower order terms}.\nonumber
\end{equation}
From the second-order determining equations of the Poisson brackets of the three integrals $X_1$, $X_2$ and $Y_3$ with the Hamiltonian $H$, we can take the compatibility conditions for the linear terms of the integrals of motion to get an overdetermined system of linear partial differential equations for the magnetic field, which can be solved. (When solving the compatibility conditions for the magnetic field, there is an additional term in the magnetic field of the form $(x^2+y^2)^{-1}$ appearing in addition to the terms present in the final result (\ref{Bgen}) below, but it vanishes either because of the compatibility conditions of the linear determining equations or the involution of the integrals $X_1$ and $X_2$, depending on the branch of the calculation.) Using the solution of the magnetic field from the compatibility conditions, one can solve the second-order determining equations to get the linear coefficients in the momenta. Then, by considering the compatibility conditions of the linear determining equations together with the zeroth-order determining equations, it is possible to find the potential $W$. By requiring that $X_1$ and $X_2$ are in involution, we finally arrive at a single system described in (\ref{Hgen}) below. Notice that the integrals $X_2$ and $Y_3$ turn out to be in involution even if it was not required. The new quadratically superintegrable system is characterized by the Hamiltonian
\begin{eqnarray}
H&=&\frac{(p_x^A)^2+(p_y^A)^2+(p_z^A)^2}{2}-\frac{r^2}{32}\left(2b_z-4b_lz+b_q(r^2+4z^2)\right)^2\nonumber\\
& &+u_1z+\frac{u_2}{r^2}+u_3(r^2+4z^2)\label{Hgen}
\end{eqnarray}
together with the magnetic field
\begin{eqnarray}
B&=&(b_lx-2xzb_q)\, dy\wedge dz+(b_ly-2b_qyz)\, dz\wedge dx\nonumber\\
& &+\left(b_z-2b_lz+b_q(r^2+2z^2)\right)dx\wedge dy,\label{Bgen}
\end{eqnarray}
where $r$ is the cylindrical radius, i.e. $r^2=x^2+y^2$. The corresponding vector potential can be chosen as
\begin{eqnarray}
\hspace{-10mm}A=-\left(\frac{b_z}{2}-b_lz+\frac{b_q}{4}(r^2+4z^2)\right)y dx+ \left(\frac{b_z}{2}-b_lz+\frac{b_q}{4}(r^2+4z^2)\right)x dy.
\end{eqnarray}
The constants $u_i$ only appear in the potential while the constants $b_i$ also appear in the magnetic field.

In the case where $b_q$ is not zero, it is possible to use a translation in $z$,
\begin{eqnarray}
z\rightarrow z+\frac{b_l}{2b_q},
\end{eqnarray}
to eliminate the constant $b_l$ from the system. After redefining the parameters $u_1$ and $b_z$, this system reads
\begin{eqnarray}
\hspace{-23mm}H&=&\frac{(p_x^A)^2+(p_y^A)^2+(p_z^A)^2}{2}-\frac{r^2}{32}\left(2b_z+b_q(r^2+4z^2)\right)^2+u_1z+\frac{u_2}{r^2}+u_3(r^2+4z^2) \label{CPHmin}
\end{eqnarray}
together with the magnetic field
\begin{eqnarray}
\hspace{-10mm}B&=&-2b_qxz\, dy\wedge dz-2b_qyz\, dz\wedge dx+\left(b_z+b_q(r^2+2z^2)\right)dx\wedge dy.
\end{eqnarray}
The vector potential $A$ can be chosen as
\begin{equation}
A=-\left(\frac{b_z}{2}+\frac{b_q}{4}(r^2+4z^2)\right)y dx+ \left(\frac{b_z}{2}+\frac{b_q}{4}(r^2+4z^2)\right)x dy.\label{ACP}
\end{equation}
The integrals of motion are given by
\begin{eqnarray}
\hspace{-10mm}X_1&=&L_x^Ap_y^A-L_y^Ap_x^A+\left(b_z+b_q(r^2+2z^2)\right)zL_z^A-\frac{b_z^2}{4}zr^2-\frac{b_zb_q}{2}zr^2(r^2+2z^2)\nonumber\\
\hspace{-10mm}& &-\frac{b_q^2}{16}zr^2(3r^2+4z^2)(r^2+4z^2)+\frac{u_1}{2}r^2-\frac{2u_2z}{r^2}+2u_3zr^2,\\
\hspace{-10mm}X_2&=&L_z^A-\left(\frac{b_z}{2}+\frac{b_q}{4}(r^2+4z^2)\right)r^2,\\
\hspace{-10mm}Y_3&=&(p_z^A)^2+2b_qz^2L_z^A-\left(b_zb_q+\frac{b_q^2}{2}(r^2+4z^2)\right)z^2r^2+2u_1z+8u_3z^2.
\end{eqnarray}
The Poisson bracket $\lbrace X_1,Y_3\rbrace$ is the only one that is not zero, but the algebra closes polynomially, i.e. the Poisson bracket squared can be expressed as a polynomial in terms of $H$, $X_1$, $X_2$ and $Y_3$,
\begin{eqnarray}
\hspace{-10mm}(\lbrace Y_3,X_1\rbrace)^2&=&-16H^2Y_3+16HY_3^2- 4Y_3^3+ 8(b_qX_2^2Y_3 + 2b_zHY_3\nonumber\\
& & + b_qX_1^2 - b_zY_3^2)X_2- 16u_1HX_1 + 32u_3X_1^2 + 8u_1X_1Y_3 \nonumber\\
& &+ 4(8u_3-b_z^2)X_2^2Y_3+ 8( b_zu_1 X_1 + 2b_qu_2Y_3)X_2\nonumber\\
& &+4u_1^2X_2^2 + 64u_2u_3Y_3 + 8u_1^2u_2.
\end{eqnarray}

The associated equations of motion in the cylindrical coordinates,
\begin{eqnarray}
\dot{r}=p_r,\qquad \dot{p_r}=\frac{p_\theta^2 + 2u_2}{r^3}-\frac{b_q p_\theta + 4u_3}{2}r,\nonumber\\
\dot{z}=p_z,\qquad \dot{p_z}=- (8u_3 +2b_qp_\theta)z  - u_1,\label{eqmothetaCP}\\
\dot{\theta}=\frac{b_q}{4}r^2+\frac{p_\theta}{r^2}+b_q z^2+\frac{b_z}{2},\qquad \dot{p_\theta}=0,\nonumber
\end{eqnarray}
can be solved,
\begin{eqnarray}
r(t)&=&\sqrt{c_1\cos(\nu t+c_2)+\sqrt{c_1^2+4\frac{L_z^2+2u_2}{\nu^2}}},\label{solrCP}\\
z(t)&=&c_3\cos(\nu t+c_4)-\frac{u_1}{\nu^2},\label{solzCP}\\
\theta(t)&=&c_5+k_1t+\frac{L_z}{\sqrt{L_z^2+2u_2}}\arctan\left(k_2\tan\left(\frac{\nu t+c_2}{2}\right)\right)\nonumber\\
& &+\frac{b_q}{\nu}\left(\frac{c_1}{4}\sin(\nu t+c_2)-\frac{2c_3u_1}{\nu^2}\sin(\nu t+c_4)\right.\nonumber\\
& &\left.+\frac{c_3^2}{2}\cos(\nu t+c_4)\sin(\nu t+c_4)\right),\label{solthetaCP}
\end{eqnarray}
where the $c_i$ and $L_z$ are integration constants and the constants $k_i$ are given by
\begin{eqnarray}
k_1&=&\frac{b_z}{2}+b_q\left(\frac{c_3^2}{2}+\frac{ u_1^2}{\nu^4}+\frac{1}{4}\sqrt{c_1^2+4\frac{L_z^2+2u_2}{\nu^2}}\right),\\
k_2&=&\sqrt{1+\frac{c_1^2\nu^2}{4(L_z^2+2u_2)}}-\frac{c_1\nu}{2\sqrt{L_z^2+2u_2}}.
\end{eqnarray}
The frequency of $r$ and $z$ is determined by the (initial) angular momentum $L_z$ and the constants $u_3$ and $b_q$, i.e. 
\begin{equation}
\nu=\sqrt{8u_3+2b_qL_z}.
\end{equation}
We can see that also this system does not possess the behaviour of a maximally superintegrable system unless there are some restrictions on $k_1$, which are not satisfied for all initial data. Hence, we conclude that for generic values of the parameters the system is minimally superintegrable. However, particular superintegrability can appear when $\theta$ matches the frequency of $r$ and $z$ up to a multiplication by a rational number.

In the case where $b_q=0$, $b_l\neq0$, we can use a different translation in $z$ to absorb the constant magnetic field, i.e.
\begin{eqnarray}
z\rightarrow z+\frac{b_z}{2b_l},
\end{eqnarray}
such that the Hamiltonian becomes 
\begin{eqnarray}
H=\frac{(p_x^A)^2+(p_y^A)^2+(p_z^A)^2}{2}-\frac{b_l^2}{2}z^2r^2+u_1z+\frac{u_2}{r^2}+u_3(r^2+4z^2)
\end{eqnarray}
(up to a redefinition of the parameter $u_1$) together with the magnetic field
\begin{eqnarray}
B=b_l\left(x\, dy\wedge dz +y\, dz\wedge dx-2 z\, dx\wedge dy\right).
\end{eqnarray}
The corresponding vector potential can be chosen as
\begin{eqnarray}
A=b_l yz dx-b_lxz dy.
\end{eqnarray}
The three integrals of motion of this system can be written as
\begin{eqnarray}
\hspace{-25mm}X_1&=&L_x^Ap_y^A-L_y^Ap_x^A-\frac{b_l}{2}(r^2+4z^2)L_z^A+\frac{u_1}{2}r^2-\frac{2u_2z}{r^2}+2u_3zr^2-\frac{b_l^2}{2}zr^2(r^2+2z^2),\\
\hspace{-25mm}X_2&=&L_z^A+b_lzr^2,\\
\hspace{-25mm}Y_3&=&(p_z^A)^2-2b_lzL_z^A+2u_1z+8u_3z^2-2b_l^2z^2r^2.
\end{eqnarray}
In the cylindrical coordinates, we can solve the associated equations of motion,
\begin{eqnarray}
r(t)&=&\sqrt{c_1 \cos(\sqrt{8u_3}t + c_2) + \sqrt{c_1^2+\frac{L_z^2 + 2u_2}{2u_3}}},\\
z(t)&=&c_3 \cos(\sqrt{8u_3}t + c_4) +\frac{b_lL_z-u_1}{8u_3},\\
\theta(t)&=&c_5-\frac{b_l(b_lL_z-u_1)}{8u_3}t-\frac{c_3b_l}{\sqrt{8u_3}}\sin(\sqrt{8u_3}t+c_4)\\
& &+\frac{L_z}{\sqrt{L_z^2+2u_2}}\arctan\left(k\tan\left(\frac{\sqrt{8u_3}t+c_2}{2}\right)\right),
\end{eqnarray}
where
\begin{eqnarray}
k=\sqrt{1+\frac{2c_1^2u_3}{L_z^2+2u_2}}-c_1\sqrt{\frac{2u_3}{L_z^2+2u_2}}.
\end{eqnarray}
Once again, we can see that this system does not possess the quality of maximally superintegrable systems in general, i.e. is minimally superintegrable.

When $b_q=b_l=0$, we can absorb the parameter $u_1$ using a translation of $z$ if $u_3$ is non-zero. In addition, similarly to the previous section, if we set $u_2=0$ and $u_3=\frac{n^2b_z^2}{8m^2}$ where $n$ and $m$ are integers, the time-dependent rotation (\ref{rotFrame5}) maps this system to a harmonic oscillator without the magnetic field, of the form
\begin{eqnarray}
H=\frac{p_x^2+p_y^2+p_z^2}{2} + \frac{n^2b_z^2}{8 m^2}(x^2+y^2+4 z^2),
\end{eqnarray}
which is known to be maximally superintegrable, cf. \cite{Evans90}. (The elimination of $u_2$, i.e. of the term $\frac{u_2}{r^2}$ in the Hamiltonian, is needed for the maximal superintegrability of the system without magnetic field, while the relation between $u_3$ and $b_z$ ensures that the frequencies of all variables match modulo integers.) We can map back its integrals of motion (one of which we obtain from the ladder operator following \cite{Marquette12}) and eliminate the time dependence arising through the rotation by taking a suitable combination of them analogously to \cite{EV08} -- that is possible under the rationality condition relating $u_3$ and $b_z$. Thus we obtain an isochronous maximally superintegrable system. Its Hamiltonian reads
\begin{eqnarray}
H=\frac{(p_x^A)^2+(p_y^A)^2+(p_z^A)^2}{2}+\frac{b_z^2}{8}\left(\frac{n^2}{m^2}-1\right)(x^2+y^2)+\frac{n^2b_z^2}{2m^2}z^2\label{MaxHam6}
\end{eqnarray}
with a constant magnetic field of magnitude $b_z$ oriented along the $z$-axis. The integrals of motion are
\begin{eqnarray}
X_1&=&L_x^Ap_y^A-L_y^Ap_x^A+b_z z L_z^A+\frac{b_z^2}{4}\left(\frac{n^2}{m^2}-1\right)zr^2,\\
X_2&=&L_z^A-\frac{b_z}{2}r^2,\\
Y_3&=&(p_z^A)^2+\frac{n^2b_z^2}{m^2}z^2,\\
Y_4&=&\mbox{Re}\left(\left((n^2 - m^2)b_z^2(y+{\rm i}x)^2- 4m^2b_z(x-{\rm i}y)(p_y^A+{\rm i}p_x^A )\right.\right.\nonumber\\
& & \left.\left. + 4m^2(p_y^A +{\rm i}p_x^A)^2\right)^n(m p_z^A+{\rm i} n b_z z)^m\right),
\end{eqnarray}
The integral $Y_4$ is of order $2 n +m$ in momenta and can be expressed without using complex expressions (as in section \ref{SecQuadOP} or \cite{EV08,MS18}). As an example, in the case $m=n=1$, the integral $Y_4$ becomes up to a rescaling
\begin{eqnarray}
Y_4=p_z^A\left((p_x^A)^2-(p_y^A)^2+b_z(xp_y^A+yp_x^A)\right)+2b_zzp_x^Ap_y^A-b_z^2z(xp_x^A-yp_y^A).\nonumber
\end{eqnarray}

Once again, this construction of maximally superintegrable systems is not exhaustive. We used the systems without magnetic fields to generate new systems admitting magnetic fields. This procedure doesn't exclude the hypothetical possibility of some other cases of maximal superintegrability belonging only to systems with magnetic fields.

\section{Conclusions}\label{SecConc}\setcounter{equation}{0}
In conclusion, we continued to investigate the 3D non-subgroup-type integrable systems that admit non-zero magnetic fields and an axial symmetry. More precisely, we looked for additional linear integrals of motion for the oblate and prolate spheroidal cases in a general manner. We found that there are only two such superintegrable systems that admit magnetic fields. These two systems were already known in the literature. We also searched for quadratically superintegrable systems in the oblate spheroidal case, in the prolate spheroidal case and in the circular parabolic case under the assumption that a well-defined limit of the Hamiltonian and the integrals of motion exists as $\vec{B}\rightarrow0$ (and the integrals remain functionally independent). We found a new quadratically minimally superintegrable system lying at the intersection of the oblate spheroidal case, the prolate spheroidal case, the cylindrical case and the spherical case. In addition, we found a new quadratically minimally superintegrable system lying at the intersection of the circular parabolic case and the cylindrical case. For both quadratically minimally superintegrable systems, we were able to solve the equations of motion and from their structure we can see that these systems cannot be maximally superintegrable in their general forms. However, with additional conditions on these systems, which make them isochronous, we were able to find two infinite families of maximally superintegrable systems. These maximally superintegrable systems involving a constant magnetic field along the $z$-axis are linked to the harmonic and caged oscillators without magnetic fields, respectively, via a rotating-frame transformation. We notice that superintegrability depends on a delicate interplay among the parameters specifying the magnetic field and the potential.

This research can be extended in many directions. The quantum versions of these systems were not studied. We only know from [1] and explicit calculations that the three integrable cases considered here (and their linear superintegrable cases) do not have quantum corrections. It would also be interesting to look in a general way for all additional quadratic integrals of motion. However, these calculations are tremendous. New techniques for finding higher-order integrals would be of great help in this matter. It would also be interesting to study the other non-subgroup-type integrable systems and then look for superintegrability.

\section*{Acknowledgements}
The research was partially supported by the Czech Science Foundation (GACR), project 17-11805S. SB was partially supported by postdoctoral fellowships provided by the Fonds de Recherche du Qu\'ebec : Nature et Technologie (FRQNT) and by the Natural Sciences and Engineering Research Council of Canada (NSERC). OK was partially supported by the Grant Agency of the Czech Technical University in Prague, grant No. SGS19/183/ OHK4/3T/14. The authors thank Antonella Marchesiello (CTU) and Pavel Winternitz (Universit\'e de Montr\'eal) for interesting discussions on the subject of this paper.

\pagebreak

\section*{Appendix --- Examples of trajectories}\label{SecFig}\setcounter{equation}{0}

\begin{figure}[h!]
\centering
\caption{Trajectory of the minimally superintegrable system (\ref{OPHmin}) with the initial values $[x(0)=1$, $y(0)=-1$, $z(0)=1$, $p_x(0)=1$, $p_y(0)=0$, $p_z(0)=0$] and the values of the constants [$u_1=2$, $u_2=\frac{3}{2}$, $u_3=-1$, $b_z=7$, $b_p=4$, $b_s=2$]. The trajectory is red at $t=0$ and becomes gradually blue.}
\begin{subfigure}{.49\textwidth}
  \centering
  \includegraphics[width=\linewidth]{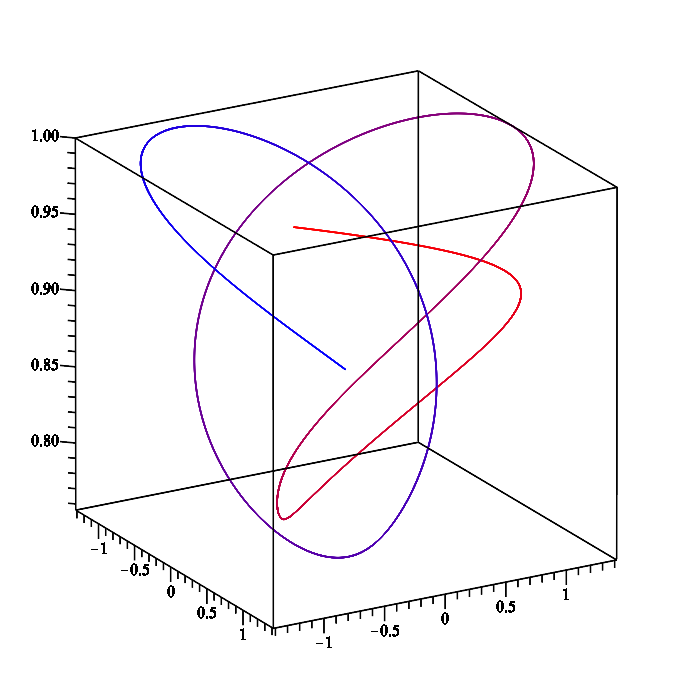}
  \caption{t=[0,2]}
\end{subfigure}
\begin{subfigure}{.49\textwidth}
  \centering
  \includegraphics[width=1.1\linewidth]{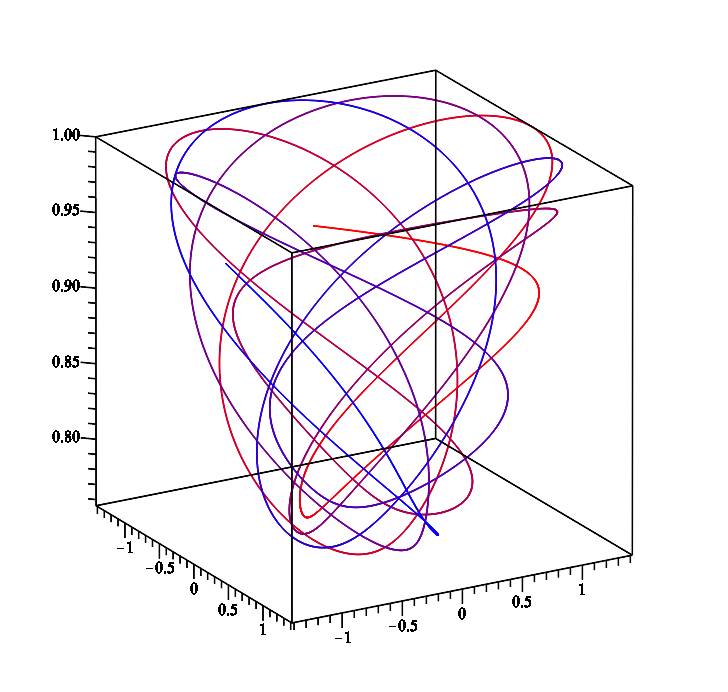}
  \caption{t=[0,7]}
\end{subfigure}
\begin{subfigure}{.49\textwidth}
  \centering
  \includegraphics[width=\linewidth]{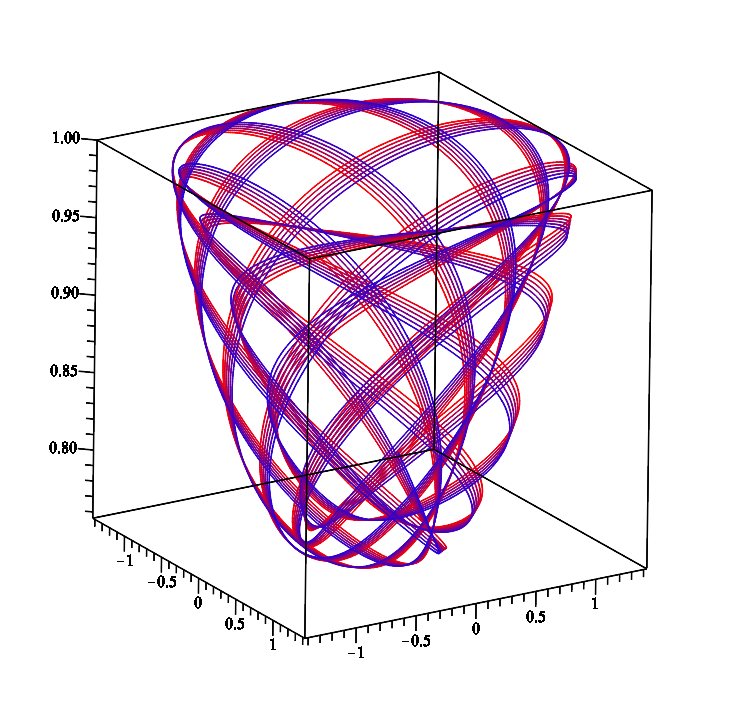}
  \caption{t=[0,50]}
\end{subfigure}
\begin{subfigure}{.49\textwidth}
  \centering
  \includegraphics[width=\linewidth]{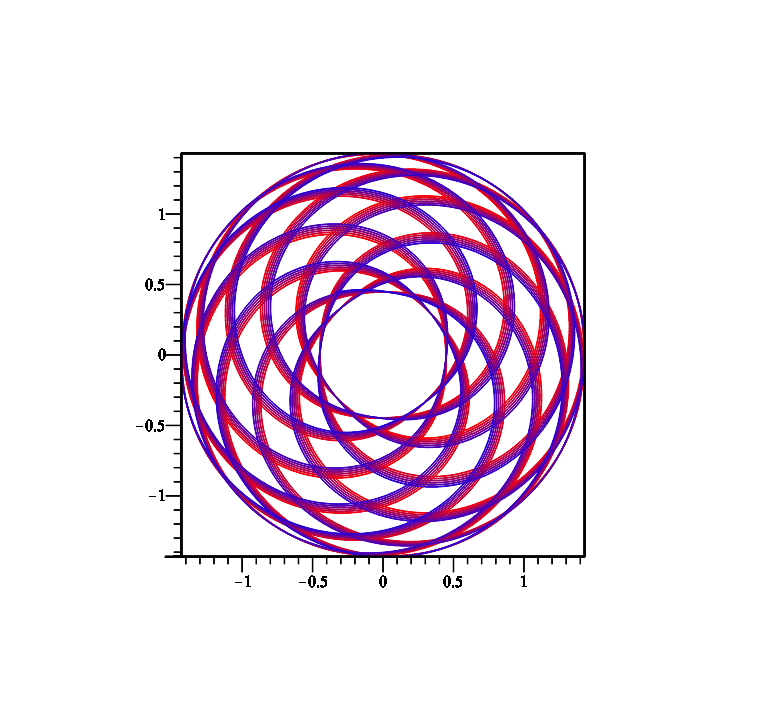}
  \caption{Projection in the xy-plane for t=[0,50]}
\end{subfigure}
\end{figure}

\begin{figure}[h!]
\centering
\caption{Trajectory of the minimally superintegrable system (\ref{OPHmin}) with the initial values $[x(0)=1$, $y(0)=-1$, $z(0)=1$, $p_x(0)=1$, $p_y(0)=0$, $p_z(0)=0$] and the values of the constants [$u_1=1$, $u_2=\frac{3}{2}$, $u_3=-1$, $b_z=2$, $b_p=4$, $b_s=2$]. The trajectory is red at $t=0$ and becomes gradually blue until it closes (around $t=18.85$).}
\begin{subfigure}{.49\textwidth}
  \centering
  \includegraphics[width=\linewidth]{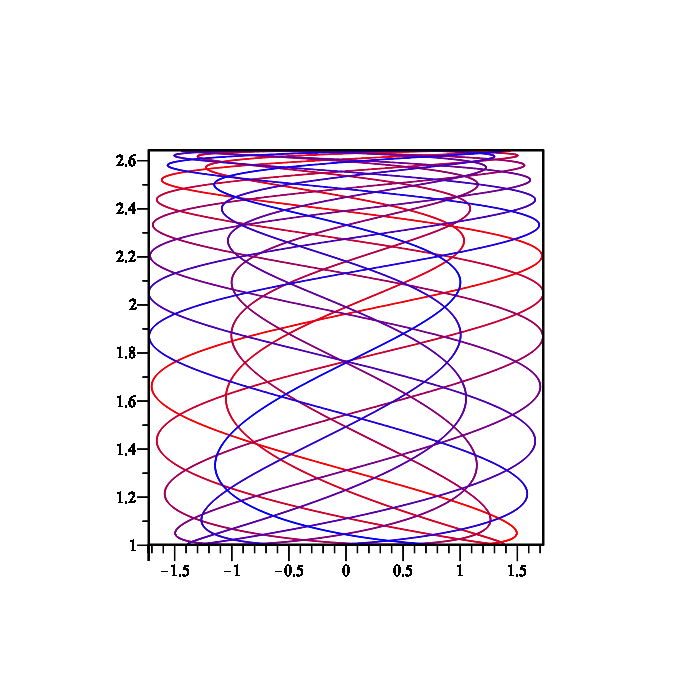}
  \caption{Projection on the xz-plane}
\end{subfigure}
\begin{subfigure}{.49\textwidth}
  \centering
  \includegraphics[width=\linewidth]{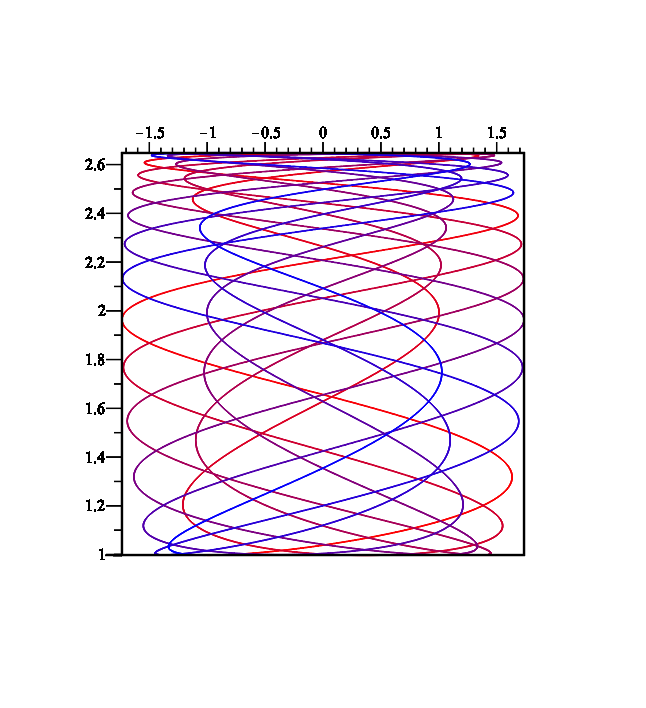}
  \caption{Projection on the yz-plane}
\end{subfigure}
\begin{subfigure}{.49\textwidth}
  \centering
  \includegraphics[width=\linewidth]{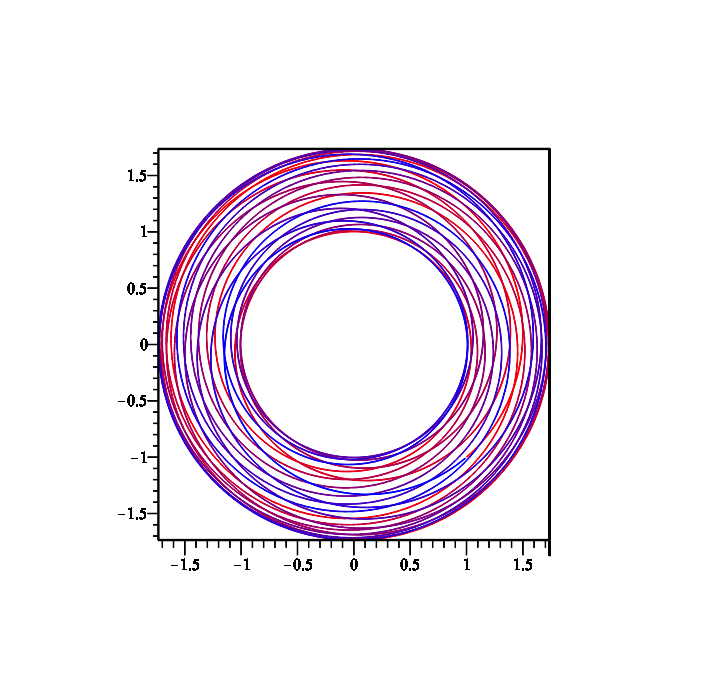}
  \caption{Projection on the xy-plane}
\end{subfigure}
\begin{subfigure}{.49\textwidth}
  \centering
  \includegraphics[width=\linewidth]{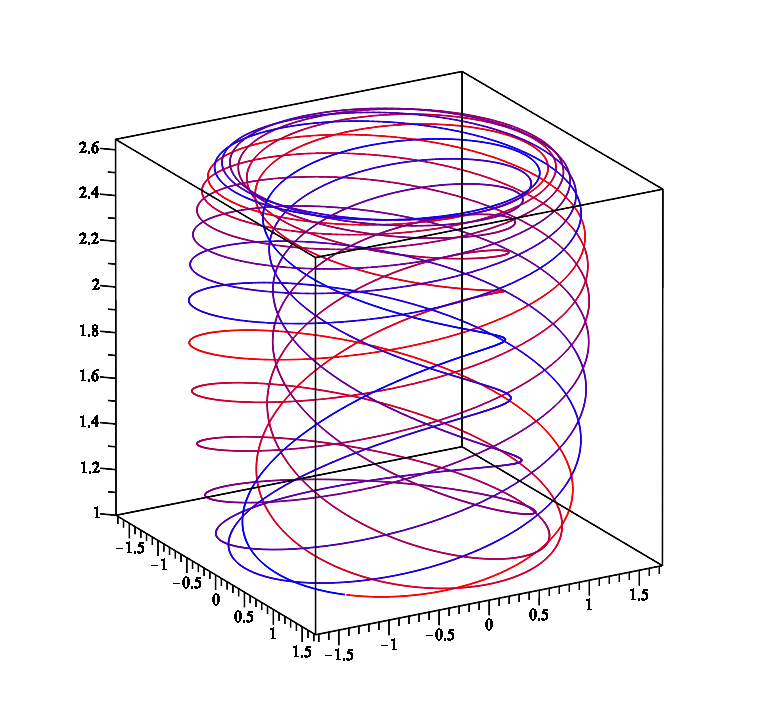}
  \caption{3D Trajectory}
\end{subfigure}
\end{figure}

\begin{figure}[h!]
\centering
\caption{Trajectory of the maximally superintegrable system (\ref{MaxHam5}) with the initial values $[x(0)=1$, $y(0)=-1$, $z(0)=1$, $p_x(0)=1$, $p_y(0)=0$, $p_z(0)=0$] and the values of the constants [$u_2=\frac{3}{2}$, $b_z=2$, $n=3$, $m=2$]. The trajectory is red at $t=0$ and becomes gradually blue until it closes  (at $t=8\pi$).}
\begin{subfigure}{.49\textwidth}
  \centering
  \includegraphics[width=\linewidth]{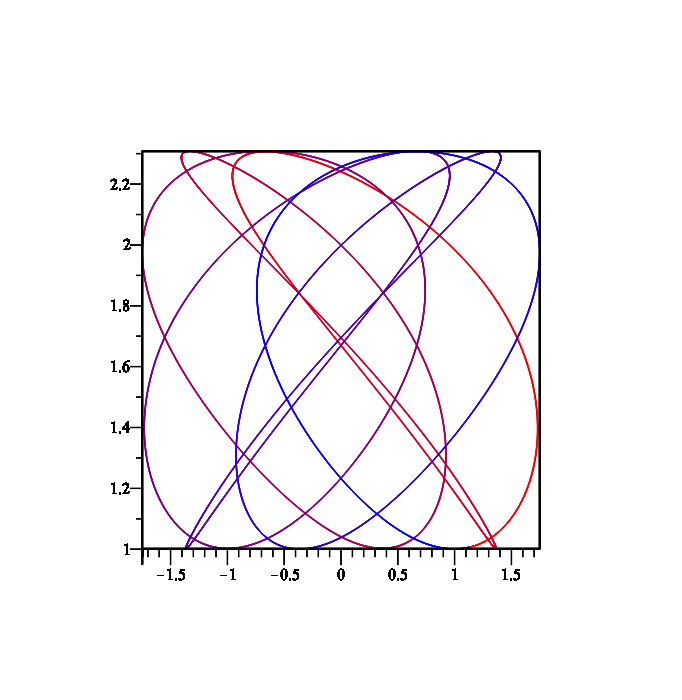}
  \caption{Projection on the xz-plane}
\end{subfigure}
\begin{subfigure}{.49\textwidth}
  \centering
  \includegraphics[width=\linewidth]{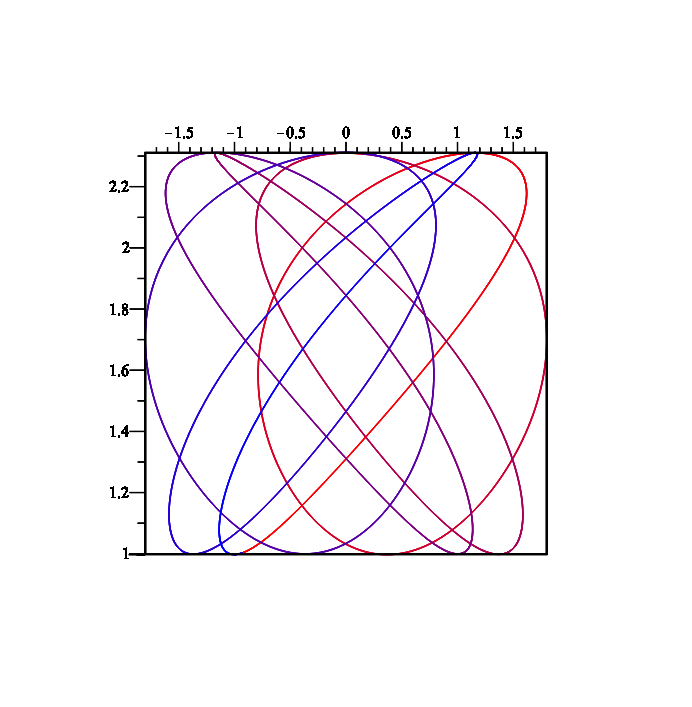}
  \caption{Projection on the yz-plane}
\end{subfigure}
\begin{subfigure}{.49\textwidth}
  \centering
  \includegraphics[width=\linewidth]{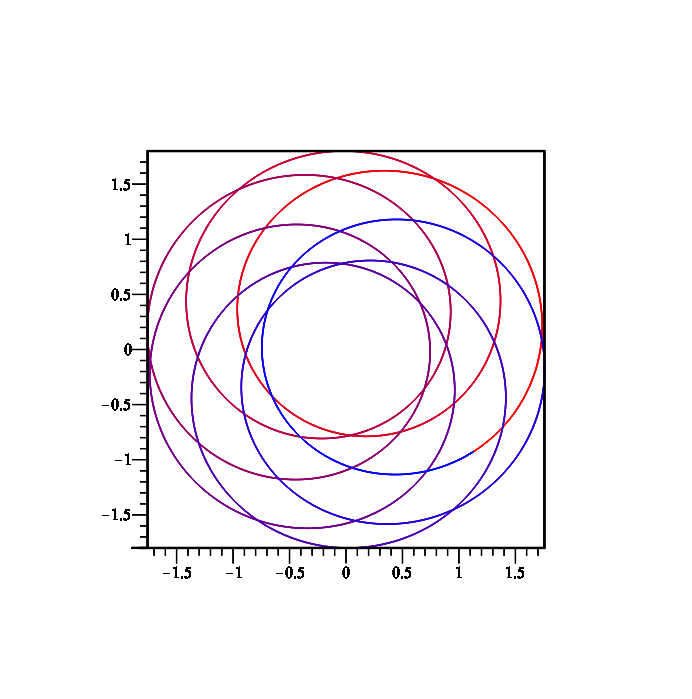}
  \caption{Projection on the xy-plane}
\end{subfigure}
\begin{subfigure}{.49\textwidth}
  \centering
  \includegraphics[width=\linewidth]{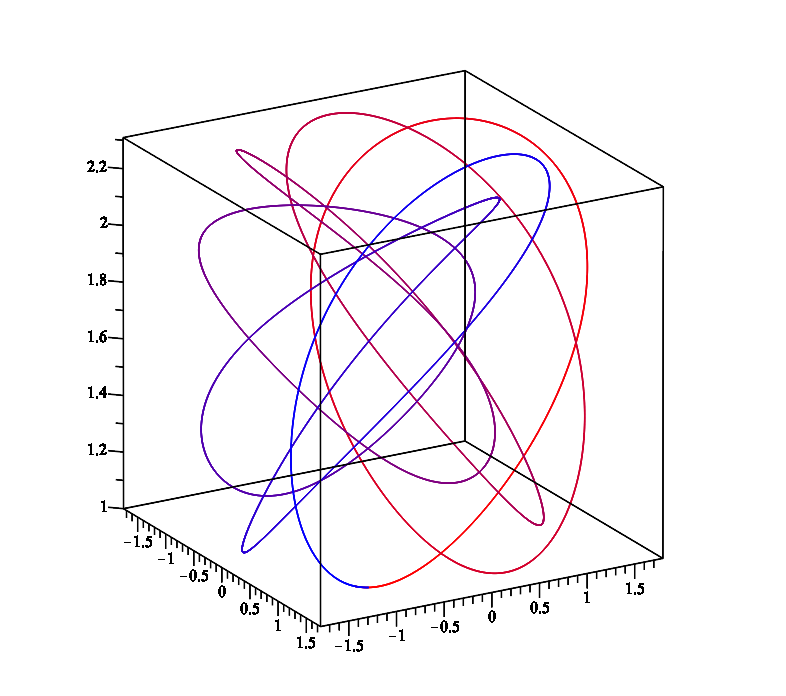}
  \caption{3D Trajectory}
\end{subfigure}
\end{figure}

\begin{figure}
\centering
\caption{Trajectory of the minimally superintegrable system (\ref{CPHmin}) with the initial values $[x(0)=1$, $y(0)=-1$, $z(0)=1$, $p_x(0)=1$, $p_y(0)=0$, $p_z(0)=0$] and the values of the constants [$u_1=10$, $u_2=\frac{3}{2}$, $u_3=1$, $b_z=2$, $b_q=4$]. The trajectory is red at $t=0$ and becomes gradually blue.}
\begin{subfigure}{.49\textwidth}
  \centering
  \includegraphics[width=\linewidth]{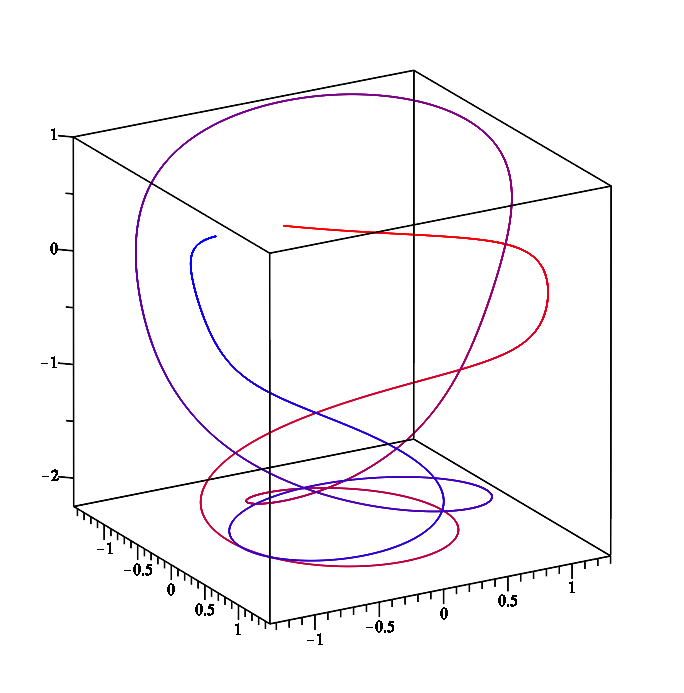}
  \caption{t=[0,3]}
\end{subfigure}
\begin{subfigure}{.49\textwidth}
  \centering
  \includegraphics[width=1.1\linewidth]{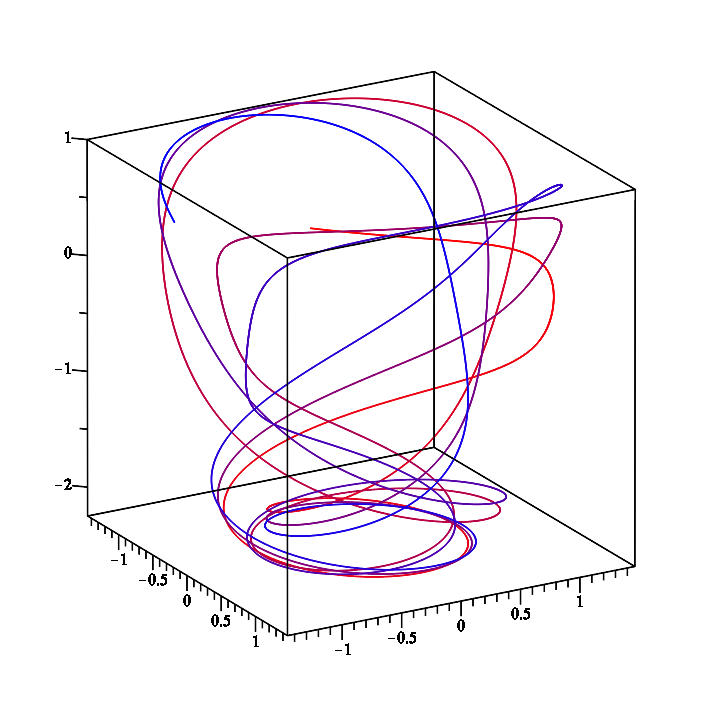}
  \caption{t=[0,8]}
\end{subfigure}
\begin{subfigure}{.49\textwidth}
  \centering
  \includegraphics[width=\linewidth]{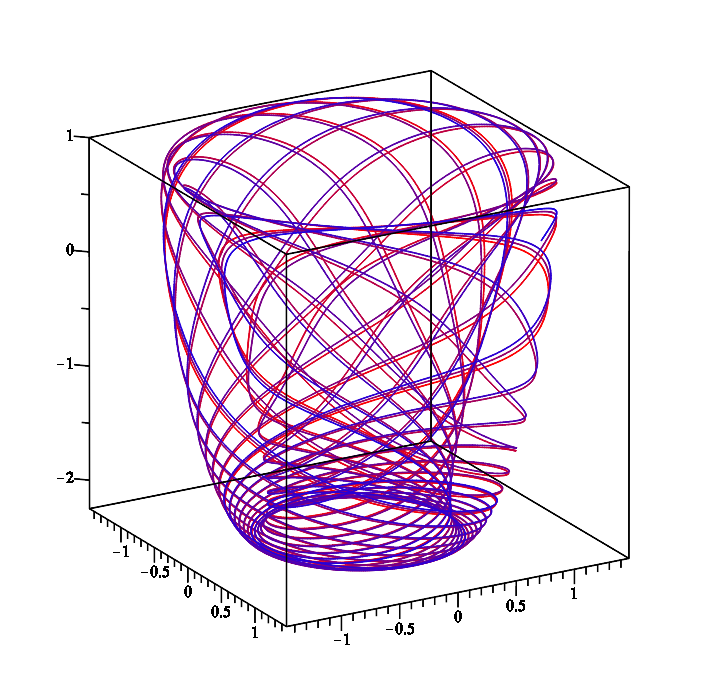}
    \caption{t=[0,50]}
\end{subfigure}
\begin{subfigure}{.49\textwidth}
  \centering
  \includegraphics[width=\linewidth]{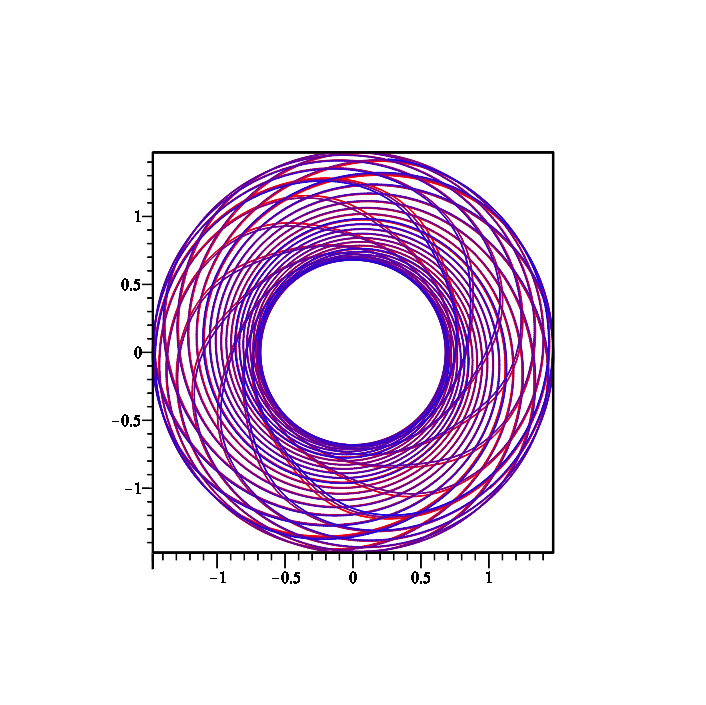}
  \caption{Projection in the xy-plane for t=[0,50]}
\end{subfigure}
\end{figure}

\begin{figure}[h!]
\centering
\caption{Trajectory of the minimally superintegrable system (\ref{CPHmin}) with the initial values $[x(0)=1$, $y(0)=-1$, $z(0)=1$, $p_x(0)=1$, $p_y(0)=0$, $p_z(0)=0$] and the values of the constants [$u_1=1$, $u_2=\frac{3}{2}$, $u_3=\frac{1}{2}$, $b_z=4$, $b_q=0$]. The trajectory is red at $t=0$ and becomes gradually blue until it closes  (around $t=12.57$).}
\begin{subfigure}{.49\textwidth}
  \centering
  \includegraphics[width=\linewidth]{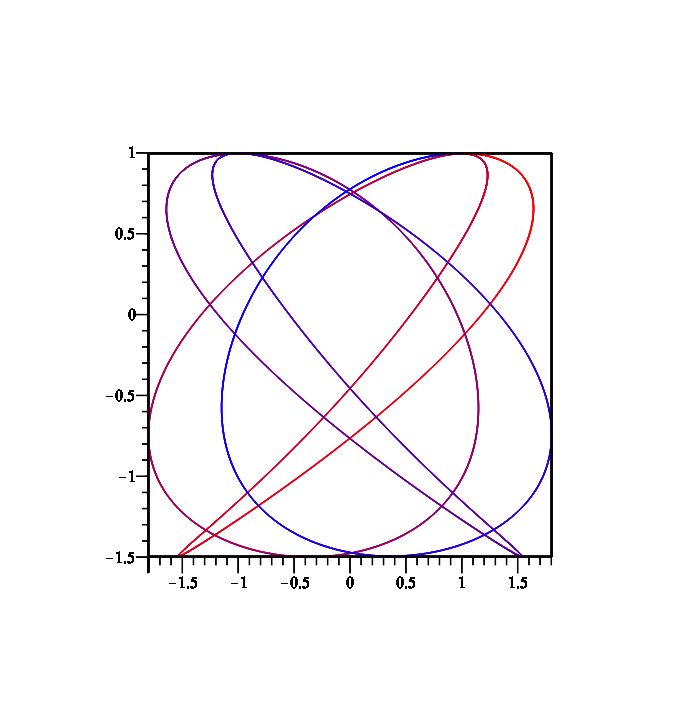}
  \caption{Projection on the xz-plane}
\end{subfigure}
\begin{subfigure}{.49\textwidth}
  \centering
  \includegraphics[width=\linewidth]{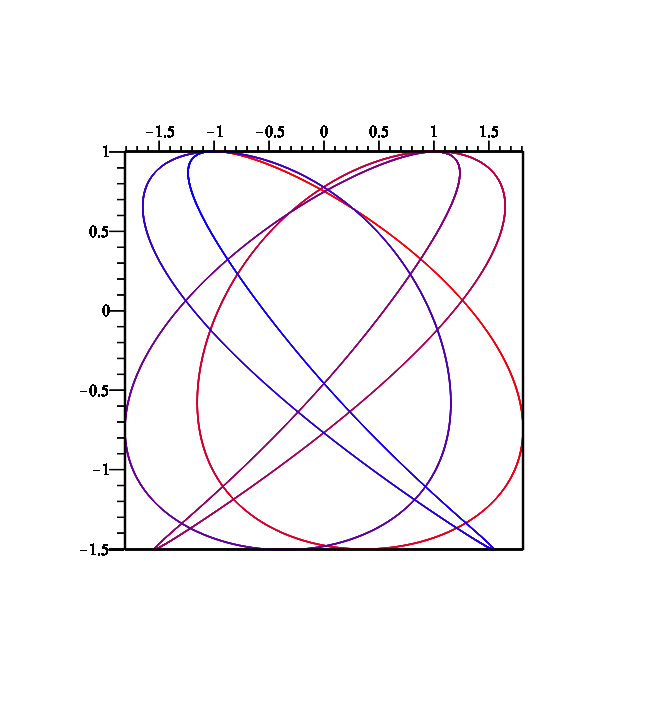}
  \caption{Projection on the yz-plane}
\end{subfigure}
\begin{subfigure}{.49\textwidth}
  \centering
  \includegraphics[width=\linewidth]{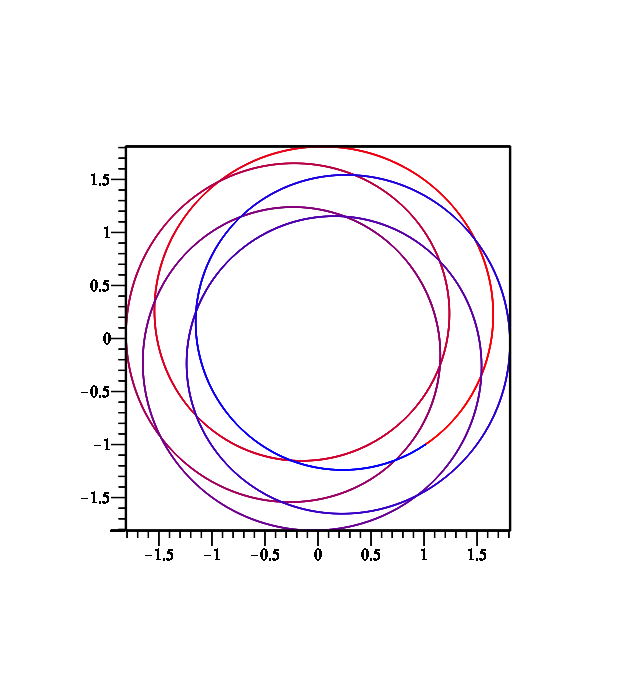}
  \caption{Projection on the xy-plane}
\end{subfigure}
\begin{subfigure}{.49\textwidth}
  \centering
  \includegraphics[width=\linewidth]{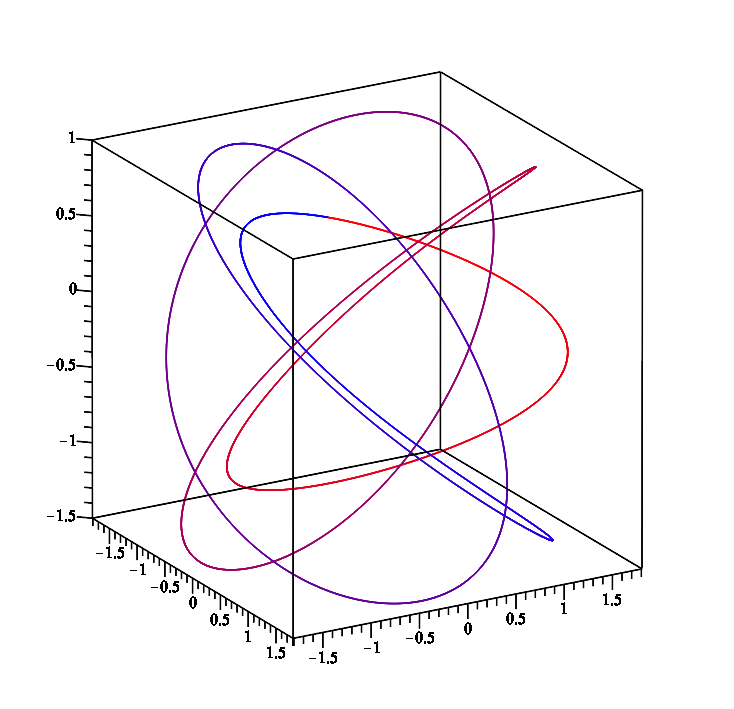}
  \caption{3D Trajectory}
\end{subfigure}
\end{figure}

\begin{figure}[h!]
\centering
\caption{Trajectory of the maximally superintegrable system (\ref{MaxHam6}) with the initial values $[x(0)=1$, $y(0)=-1$, $z(0)=1$, $p_x(0)=1$, $p_y(0)=0$, $p_z(0)=0$] and the values of the constants [$b_z=3$, $n=1$, $m=2$]. The trajectory is red at $t=0$ and becomes gradually blue until it closes  (at $t=\frac{8\pi}{3}$).}
\begin{subfigure}{.49\textwidth}
  \centering
  \includegraphics[width=\linewidth]{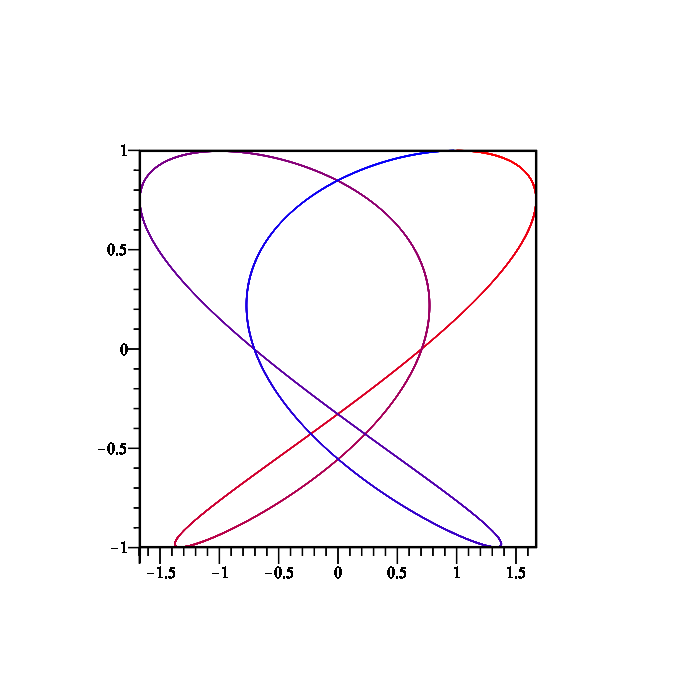}
  \caption{Projection on the xz-plane}
\end{subfigure}
\begin{subfigure}{.49\textwidth}
  \centering
  \includegraphics[width=\linewidth]{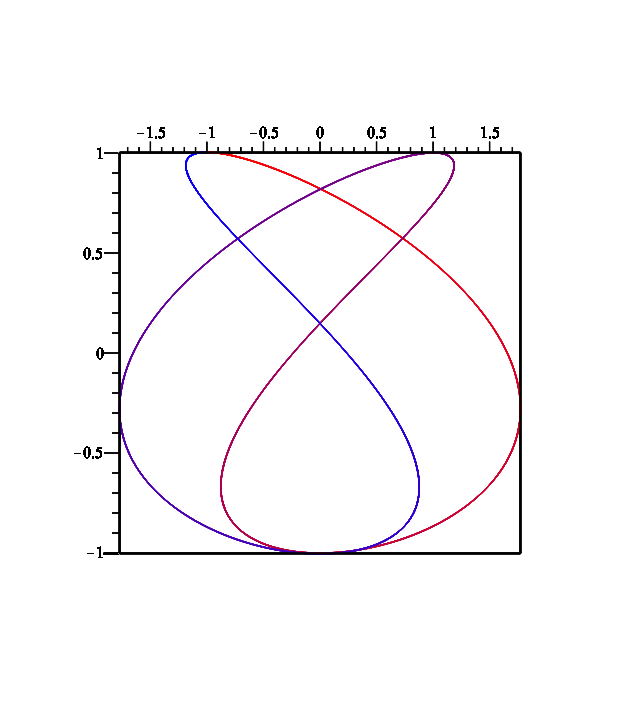}
  \caption{Projection on the yz-plane}
\end{subfigure}
\begin{subfigure}{.49\textwidth}
  \centering
  \includegraphics[width=\linewidth]{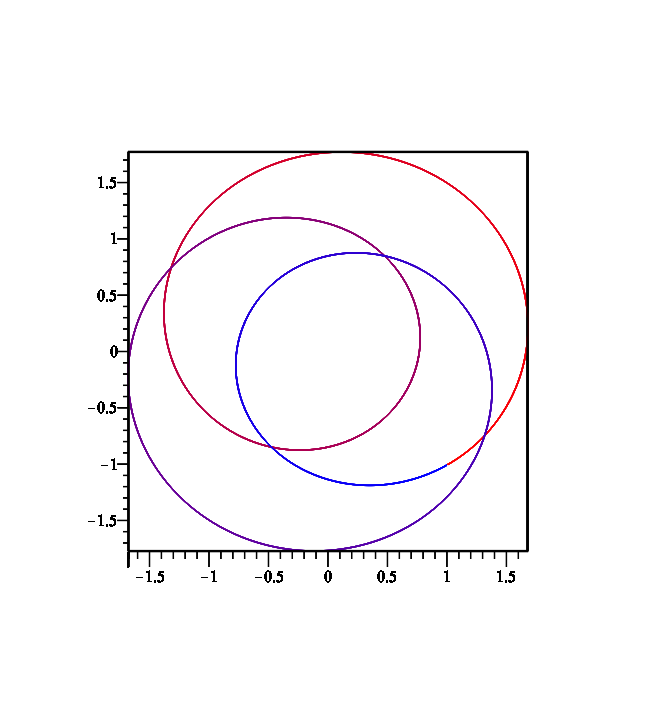}
  \caption{Projection on the xy-plane}
\end{subfigure}
\begin{subfigure}{.49\textwidth}
  \centering
  \includegraphics[width=\linewidth]{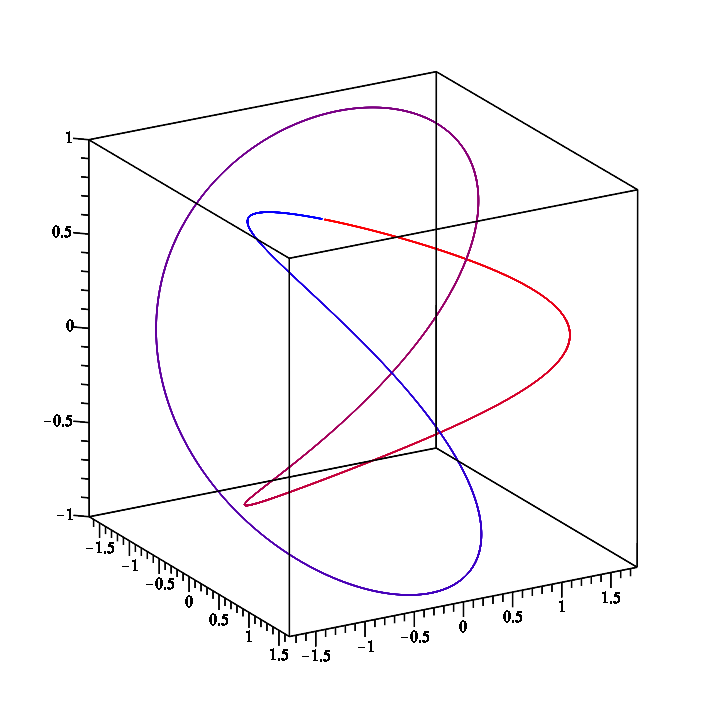}
  \caption{3D Trajectory}
\end{subfigure}
\end{figure}


\section*{References}
\begin{thebibliography}{00}

\bibitem{BS19}
Bertrand S and \v{S}nobl L (2019) On rotationally invariant integrable and superintegrable classical systems in magnetic fields with non-subgroup type integrals,\textit{ J. Phys. A: Math. Theor.} \textbf{52} 195201 (25pp). DOI: \href{http://dx.doi.org/10.1088/1751-8121/ab14c2}{10.1088/1751-8121/ab14c2}

\bibitem{BW04}
B\'erub\'e J and Winternitz P (2004) Integrable and superintegrable quantum systems in a magnetic field, \textit{J. Math. Phys} \textbf{45} 1959--1973. DOI: \href{http://dx.doi.org/10.1063/1.1695447}{10.1063/1.1695447}

\bibitem{Calogero08}
Calogero F (2008) \textit{Isochronous Systems} (Oxford University Press, Oxford)

\bibitem{DGRW}
Dorizzi B, Grammaticos B, Ramani A and Winternitz P (1985) Integrable {H}amiltonian systems with velocity-dependent potentials, \textit{J. Math. Phys.} \textbf{26} 3070--3079. DOI: \href{http://dx.doi.org/10.1063/1.526685}{10.1063/1.526685}

\bibitem{EisenhartAnnMath}
Eisenhart LP (1934) Separable systems of {S}t\"ackel, \textit{Ann. of Math.} \textbf{35} 284--305.\\
DOI: \href{http://dx.doi.org/10.2307/1968433}{10.2307/1968433}

\bibitem{ELW17}
Escobar-Ruiz AM, L\'opez Vieyra JC, Winternitz P and Yurdusen I (2018) Fourth order superintegrable systems separating in polar coordinates. {II}. {S}tandard potentials.
\newblock {\em J. Phys. A: Math. Theor.} \textbf{51} 455202. DOI: \href{http://dx.doi.org/10.1088/1751-8121/aae291}{10.1088/1751-8121/aae291}

\bibitem{Evans90}
Evans NW (1990) Superintegrability in classical mechanics, \textit{Phys. Rev. A} \textbf{41} 5666--5676. \\DOI: \href{http://dx.doi.org/10.1103/PhysRevA.41.5666}{10.1103/PhysRevA.41.5666}

\bibitem{EV08}
Evans NW and Verrier PE (2008) Superintegrability of the caged anisotropic oscillator, \textit{J. Math. Phys.} \textbf{49} 092902. DOI: \href{http://dx.doi.org/10.1063/1.2988133}{10.1063/1.2988133}

\bibitem{FSW19}
Fournier F, \v{S}nobl L and Winternitz P (2020) Cylindrical type integrable classical systems in a magnetic field, \textit{J. Phys. A: Math. Theor.} \textbf{53} 085203 (31pp). DOI: \href{http://dx.doi.org/10.1088/1751-8121/ab64a6}{10.1088/1751-8121/ab64a6}

\bibitem{FMSUW}
Fri\v{s} J, Mandrosov V, Smorodinsky YA, Uhl\'i\v{r} M and Winternitz P (1965) On higher symmetries in quantum mechanics, \textit{Phys. Lett.} \textbf{16} 354--356. DOI: \href{http://dx.doi.org/10.1016/0031-9163(65)90885-1}{10.1016/0031-9163(65)90885-1}

\bibitem{FSUW}
Fri\v{s} J, Smorodinsky YA, Uhl\'i\v{r} M and Winternitz P (1966) Symmetry groups in classical and quantum mechanics, \textit{Yad Fiz} \textbf{4} 625–-635 (1966 Sov. J. Nucl. Phys. \textbf{4} 444--450).

\bibitem{Gravel04}
Gravel S (2004) Hamiltonians separable in {C}artesian coordinates and third-order integrals of motion, \textit{ J. Math. Phys.}, \textbf{45} 1003--1019. DOI: \href{http://dx.doi.org/10.1063/1.1633352}{10.1063/1.1633352}

\bibitem{GL18}
Gubbiotti G and Latini D (2018) A multiple scales approach to maximal superintegrability, \textit{J. Phys. A: Math. Theor.} \textbf{51} 285201. DOI: \href{https://doi.org/10.1088/1751-8121/aac036}{10.1088/1751-8121/aac036}

\bibitem{KKM05}
Kalnins EG, Kress JM and Miller W~Jr (2005) Second order superintegrable systems in conformally flat spaces. {III}. {T}hree-dimensional classical structure theory. \textit{J. Math. Phys.} \textbf{46} 103507.\\ DOI: \href{http://dx.doi.org/10.1063/1.2037567}{10.1063/1.2037567}

\bibitem{KKM07}
Kalnins EG, Kress JM and Miller W~Jr (2007) Fine structure for 3{D} second-order superintegrable systems: three-parameter potentials, \textit{ J. Phys. A: Math. Theor.} \textbf{40} 5875--5892.\\ DOI: \href{http://dx.doi.org/10.1088/1751-8113/40/22/008}{10.1088/1751-8113/40/22/008}

\bibitem{KKM07o}
Kalnins EG, Kress JM, and Miller W~Jr (2007) Nondegenerate three-dimensional complex {E}uclidean superintegrable systems and algebraic varieties. \textit{J. Math. Phys.} \textbf{48} 113518.\\ DOI: \href{http://dx.doi.org/10.1063/1.2817821}{10.1063/1.2817821}

\bibitem{KKM18}
Kalnins EG, Kress JM, and Miller W~Jr (2018) \textit{Separation of variables and superintegrability}: The symmetry of solvable systems. (IOP Publishing, Bristol). DOI: \href{http://dx.doi.org/10.1088/978-0-7503-1314-8}{10.1088/978-0-7503-1314-8}

\bibitem{KWMP}
Kalnins EG, Williams GC, Miller W~Jr and Pogosyan GS (1999) Superintegrability in three-dimensional {E}uclidean space, \textit{J. Math. Phys.}, \textbf{40} 708--725. DOI: \href{http://dx.doi.org/10.1063/1.532699}{10.1063/1.532699}

\bibitem{LMV91}
Labelle S, Mayrand M and Vinet L (1991) Symmetries and degeneracies of a charged oscillator in the field of a magnetic monopole, \textit{J. Math. Phys.}, \textbf{32}1516--1521. DOI: \href{http://dx.doi.org/10.1063/1.529259}{10.1063/1.529259}

\bibitem{MSVW}
Makarov AA, Smorodinsky JA, Valiev Kh and Winternitz P (1967) A systematic search for nonrelativistic systems with dynamical symmetries, \textit{Il Nuovo Cimento A} \textbf{52} 8881--8903.\\ DOI: \href{http://dx.doi.org/10.1007/BF02755212}{10.1007/BF02755212}

\bibitem{Maple}
Maple 2019. Maplesoft, a division of Waterloo Maple Inc., Waterloo, Ontario, Canada.

\bibitem{MSW15}
Marchesiello A, \v{S}nobl L and Winternitz P (2015) Three-dimensional superintegrable systems in a static electromagnetic field,\textit{J. Phys. A: Math. Theor.} \textbf{48 }395206 (24pp).\\ DOI: \href{http://dx.doi.org/10.1088/1751-8113/48/39/395206}{10.1088/1751-8113/48/39/395206}

\bibitem{MS17}
Marchesiello A and \v{S}nobl L (2017) Superintegrable 3{D} systems in a magnetic field corresponding to {C}artesian separation of variables, \textit{J. Phys. A: Math. Theor.} \textbf{50} 245202 (24pp).\\ DOI: \href{http://dx.doi.org/10.1088/1751-8121/aa6f68}{10.1088/1751-8121/aa6f68}

\bibitem{MS18}
Marchesiello A and \v{S}nobl L (2018) An Infinite Family of Maximally Superintegrable Systems in a Magnetic Field with Higher Order Integrals, \textit{SIGMA} \textbf{14} 092 (11pp).\\ DOI: \href{http://dx.doi.org/10.3842/SIGMA.2018.092}{10.3842/SIGMA.2018.092}

\bibitem{MSW18}
Marchesiello A, \v{S}nobl L and Winternitz P (2018) Spherical type integrable classical systems in a magnetic field, \textit{J. Phys. A: Math. Theor.} \textbf{51} 135205 (24pp).  DOI: \href{http://dx.doi.org/10.1088/1751-8121/aaae9b}{10.1088/1751-8121/aaae9b}

\bibitem{MS19}
Marchesiello A and \v{S}nobl L (2020) Classical superintegrable systems in a magnetic field that separate in {C}artesian coordinates, \textit{SIGMA} \textbf{16}, 015 (35pp). DOI: \href{http://dx.doi.org/10.3842/SIGMA.2020.015}{10.3842/SIGMA.2020.015}

\bibitem{MW08}
Marquette I and Winternitz P (2008) Superintegrable systems with third-order integrals of motion, \textit{J. Phys. A: Math. Theor.} \textbf{41} 304031. DOI: \href{http://dx.doi.org/10.1088/1751-8113/41/30/304031}{10.1088/1751-8113/41/30/304031}

\bibitem{MSW17}
Marquette I, Sajedi M and Winternitz P (2017) Fourth order superintegrable systems separating in {C}artesian coordinates i. exotic quantum potentials, \textit{J. Phys. A: Math. Theor} \textbf{50} 315201.\\  DOI: \href{http://dx.doi.org/10.1088/1751-8121/aa7a67}{10.1088/1751-8121/aa7a67}

\bibitem{Marquette10}
Marquette I (2010) Superintegrability and higher order polynomial algebras, \textit{J. Phys. A: Math. Theor.} \textbf{43} 135203. DOI: \href{http://dx.doi.org/10.1088/1751-8113/43/13/135203}{10.1088/1751-8113/43/13/135203}

\bibitem{Marquette12}
Marquette I (2012) Classical ladder operators, polynomial Poisson algebras, and classification of superintegrable systems, \textit{J. Math. Phys.} \textbf{53} 012901. DOI: \href{http://dx.doi.org/10.1063/1.3676075}{10.1063/1.3676075}

\bibitem{MC10}
McIntosh HV and Cisneros A (1970) Degeneracy in the presence of a magnetic monopole, \textit{ J. Math. Phys.} \textbf{11} 896--916. DOI: \href{http://dx.doi.org/10.1063/1.1665227}{10.1063/1.1665227}

\bibitem{MW00}
McSween E and Winternitz P (2000) Integrable and superintegrable {H}amiltonian systems in magnetic fields, \textit{ J. Math. Phys.} \textbf{41} 2957--2967. DOI: \href{http://dx.doi.org/10.1063/1.533283}{10.1063/1.533283}

\bibitem{MPW13}
Miller W Jr, Post S and Winternitz P (2013) Classical and quantum superintegrability with applications, \textit{J. Phys. A: Math. Theor.} \textbf{46} 423001 (97pp).\\ DOI: \href{http://dx.doi.org/10.1088/1751-8113/46/42/423001}{10.1088/1751-8113/46/42/423001}

\bibitem{Nehorosev}
Nehoro\v{s}ev NN (1972) Action-Angle Variables and their Generalizations, \textit{Trans. Moscow Math. Soc.} \textbf{26} 180--198

\bibitem{PR05}
Pucacco G and Rosquist K (2005) Integrable {H}amiltonian systems with vector potentials, \textit{J. Math. Phys.} \textbf{46} 012701. DOI: \href{http://dx.doi.org/10.1063/1.1818721}{10.1063/1.1818721}

\bibitem{Pucacco04}
Pucacco G (2004) On integrable {H}amiltonians with velocity dependent potentials, \textit{Celestial Mechanics and Dynamical Astronomy} \textbf{90} 109--123.

\bibitem{TD11}
Tanoudis Y and Daskaloyannis C (2011) Algebraic calculation of the energy eigenvalues for the nondegenerate three-dimensional {K}epler-{C}oulomb potential. \textit{SIGMA} \textbf{7} 054.\\ DOI: \href{http://dx.doi.org/10.3842/SIGMA.2011.054}{10.3842/SIGMA.2011.054}

\bibitem{Turbiner}
Turbiner AV (2013) Particular integrability and (quasi)-exact-solvability, \textit{J. Phys A: Math. Theor.} \textbf{46} 025203. DOI: \href{http://dx.doi.org/10.1088/1751-8113/46/2/025203}{10.1088/1751-8113/46/2/025203}

\bibitem{VE08}
Verrier PE and Evans NW (2008) A new superintegrable {H}amiltonian, \textit{J. Math. Phys.} \textbf{49} 022902. DOI: \href{http://dx.doi.org/10.1063/1.2840465}{10.1063/1.2840465}

\bibitem{zhalij15}
Zhalij A (2015) Quantum integrable systems in three-dimensional magnetic fields: the Cartesian case, \textit{J. Phys.: Conf. Ser.} \textbf{621} 012019. DOI: \href{http://dx.doi.org/10.1088/1742-6596/621/1/012019}{10.1088/1742-6596/621/1/012019}

\bibitem{RotF}
Zhang PM, Zou LP, Horvathy PA and Gibbons GW (2014) Separability and dynamical symmetry of Quantum Dots, \textit{Ann. Phys.} \textbf{341} 94--116. DOI: \href{http://dx.doi.org/10.1016/j.aop.2013.11.004}{10.1016/j.aop.2013.11.004}

\end{thebibliography}
\end{document}